\documentclass[runningheads]{svmult}

\usepackage{makeidx}   
\usepackage{graphicx}  
\usepackage{subeqnar}  
\usepackage{multicol}  
\usepackage{physprbb}  
\makeindex             

\newcommand{\revised}{}

\begin{document}
\title*{MHD Turbulence in Star-Forming Regions \protect\newline and
the Interstellar Medium} 
\toctitle{MHD Turbulence in Star-Forming Regions and the Interstellar Medium}
%
%
\titlerunning{MHD Turbulence in Star-Forming Regions and ISM}
%
\author{Mordecai-Mark Mac Low\inst{1}}
\authorrunning{Mordecai-Mark Mac Low}
%
%
\institute{Dept. of Astrophysics, American Museum of Natural History,
79th Street and Central Park W., New York, NY, 10024-5192, USA;
mordecai@amnh.org}

\maketitle              

\begin{abstract}
MHD turbulence plays a central role in the physics of star-forming
molecular clouds and the interstellar medium. {\revised MHD turbulence
in molecular clouds must be driven to account for the observed
supersonic motions in the clouds, as even strongly magnetized
turbulence decays quickly.  Driven MHD turbulence can globally support
gravitationally unstable regions, but local collapse inevitably
occurs.} Differences in the strength of driving and the gas density
may explain the very different rates of star formation observed in
different galaxies.  {\revised Two types of comparisons to
observations are reviewed.  First, the use of wavelet transform
methods suggest that the driving comes from scales larger than
observed molecular clouds.  Second, comparison of simulated spectral
cubes from models to real observations suggests that Larson's
mass-size relationship is an observational artifact.} The driving
mechanism for the turbulence is likely a combination of field
supernovae in star-forming sections of galactic disks, and
magnetorotational instabilities in outer disks and low surface
brightness galaxies.  Supernova-driven turbulence has a broad range of
pressures with a roughly log-normal distribution.  High-pressure, cold
regions can be formed even in the absence of self-gravity.
\end{abstract}

\section{Introduction}

{\revised One of the big questions} in star formation is what
determines the rate of star formation in galaxies?  Another, more
pointed way of phrasing this question is to ask why the star formation
rate in normal galaxies is so low, and why it varies so strongly, over
orders of magnitude from low surface brightness galaxies, through
normal galaxies, to starburst galaxies.  The free-fall time for gas at
typical interstellar densities is
\begin{equation}
t_{\rm ff} = \sqrt{\frac{3\pi}{32 G \bar\rho}} \approx (3.4 \times 10^7
\mbox{ yr}) \left(\frac{n}{1 \mbox{ cm}^{-3}}\right)^{-1/2},
\label{equ:tff}
\end{equation}
where $\bar\rho$ is the mean mass density of the gas, $G$ the gravitational 
constant and $n=\bar\rho/\mu$ the number density, with $\mu=2.36
m_H$. Yet galactic ages range up to $10^{10}$~yr, and star formation
continues today.  What has delayed star formation sufficiently to
allow it to continue?

In what might be called the standard theory of star formation,
magnetic fields are invoked to answer both of these questions.  If
fields are strong enough, they can magnetostatically support clouds
against collapse.  The star formation rate would then be determined by
the rate of ambipolar drift of neutral gas past ions tied to the
magnetic field towards the centers of self-gravitating cores
\cite{m77,s77}.  Furthermore, if the fields are strong enough that the
Alfv\'en speed $v_A$ reaches the rms velocity $v$, then strong shocks
will be converted to MHD waves.  As linear Alfv\'en waves are
lossless, it was thought that motions remaining from the initial
formation of the clouds might be enough to explain the observation of
strongly supersonic motions in molecular clouds \cite{am75}.  In this
review I will explain why both of these ideas now appear questionable.

\section{Decaying Turbulence}

\begin{figure}[thbf]
\begin{center}
\includegraphics[width=.8\textwidth]{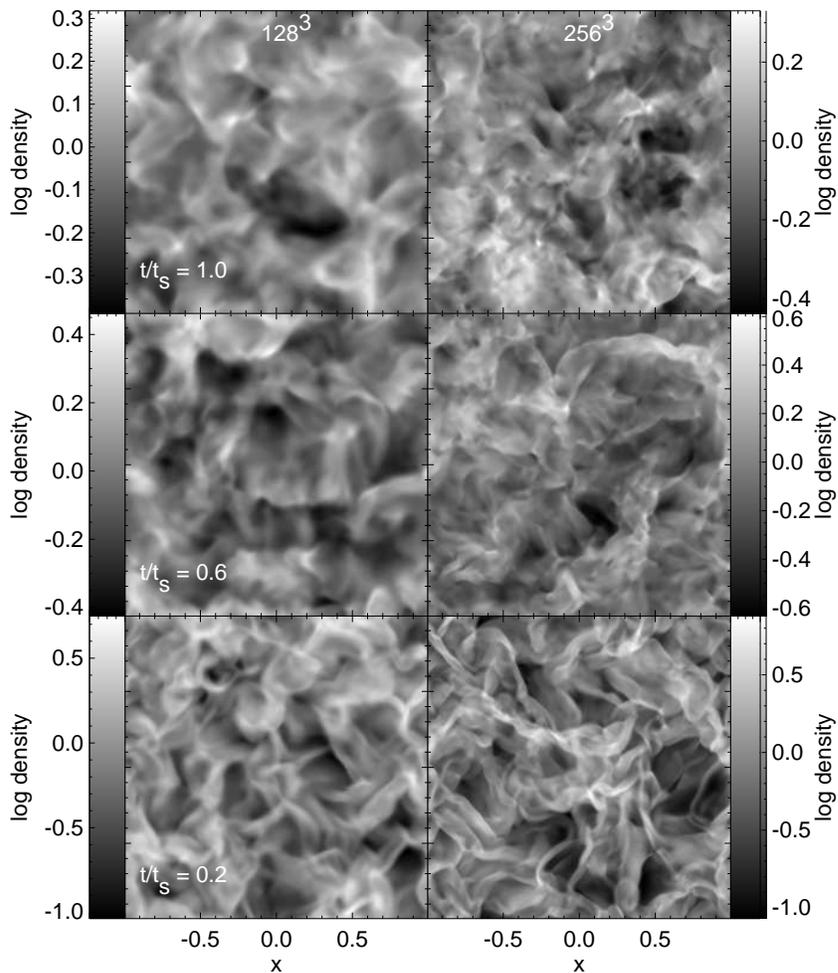}
\end{center}
\caption[Decaying isothermal turbulence morphology]{Log of density is
shown at times $t/t_s = 0.2$, 0.6 and 1.0 on slices through the
decaying supersonic hydrodynamic models C and D with initial rms Mach
number $M = 5$ described in \cite{m98} at standard resolution
($128^3$) and high resolution ($256^3$), where $t_s$ is the sound
crossing time of our numerical box.  Note that each image is scaled to
its own maximum and minimum to enhance morphological features
(From~\cite{m99})}
\label{decaysix}
\end{figure}

First let us consider the decay of supersonic turbulence as shown in
Figure~\ref{decaysix}. 
I call on computations performed with two different methods: Eulerian
hydrodynamics and MHD on a grid, using the code ZEUS-3D
\cite{sn92a,sn92b,c94,hs95}, available from the Laboratory for
Computational Astrophysics at
\begin{quote}
http://zeus.ncsa.uiuc.edu/lca\_home\_page.html,
\end{quote}
and Lagrangian hydrodynamics using a smoothed particle hydrodynamics
(SPH) code derived from that described by Benz \cite{b90} and Monaghan
\cite{m92}, running on special purpose GRAPE processors
\cite{e93,s96}, and incorporating sink particles \cite{bbp95}.

We chose initial conditions for our models inspired by the popular
idea that setting up velocity perturbations with an initial power
spectrum $P(k) \propto k^\alpha$ in Fourier space similar to that of
developed turbulence would be in some way equivalent to starting with
developed turbulence \cite{pn99,ppw92,ppw94}.  {\revised Observing the
development of our models, it became clear to us that the loss of
phase information in the power spectrum~\cite{af85} allows extremely
different gas distributions to have the same power spectrum.  This is
particularly important for supersonic flows. Supersonic,} HD
turbulence has been found in simulations \cite{ppw94} to have a power
spectrum $\alpha = -2$.  However, any single, discontinuous shock wave
will also have such a power spectrum, as that is simply the Fourier
transform of a step function, and taking the Fourier transform of many
shocks will not change this power law.  Nevertheless, most
distributions with $\alpha = -2$ do not contain shocks.

After experimentation, we decided that the quickest way to generate
fully developed turbulence was with perturbations having a flat power
spectrum $\alpha = 0$ {\revised for $0 < k_d < 8$.}
We set up velocity perturbutions drawn from a
Gaussian random field fully determined by its power spectrum in
Fourier space following the standard procedure: for each wavenumber
${\bf k_d}$ we randomly select an amplitude from a Gaussian distribution
centered on zero and with width $P(k_d)= P_{\rm 0} k_d^\alpha$ with
$k_d=|{\bf k_d}|$, and a phase between zero and $2\pi$.  We then transform
the field back into real space to obtain the velocity in each zone.
This is done independently for each velocity component.  For the SPH
calculation the velocities defined on the grid are assigned onto
individual particles using the ``cloud-in-cell'' scheme \cite{he88}.
In all of our models we take $c_{\rm s} = 0.1$, initial density
$\rho_{\rm 0} = 1$, and we use a periodic grid with sides $L = 2$
centered on the origin.  These parameter choices define our unit
system. Our choice of periodic boundary conditions corresponds to the
case of free turbulence discussed above, at least initially.
Thereafter, the appropriate treatment is less clear.

\begin{figure}[thbf]
\begin{center}
\includegraphics[width=.75\textwidth]{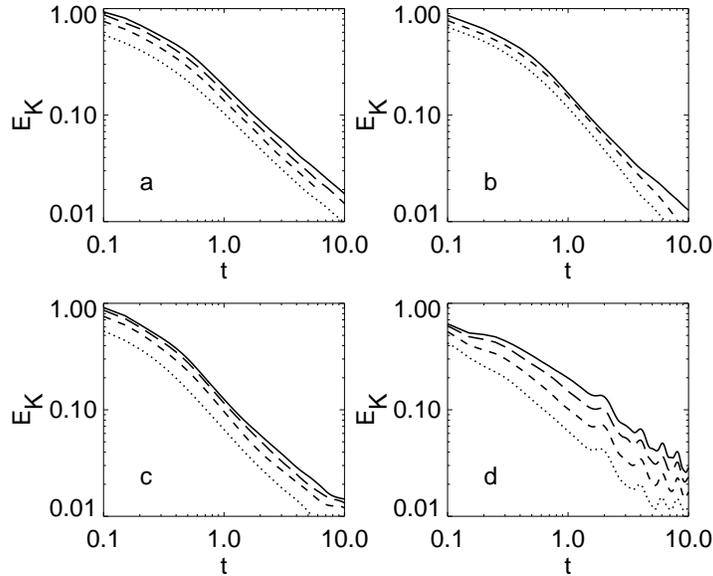}
\end{center}
\caption[Kinetic energy vs.\ time in decaying
turbulence]{Three-dimensional resolution studies for initial rms Mach
number $M=5$, isothermal models.  ZEUS models have $32^3$ ({\em
dotted}), $64^3$ ({\em short dashed}), $128^3$ ({\em long dashed}), or
$256^3$ ({\em solid}) zones, while the SPH models have 7000 ({\em
dotted}), 50,000 ({\em short dashed}), or 350,000 ({\em solid})
particles. Panels show {\em a)} hydro runs with ZEUS, {\em b)} hydro
runs with SPH, {\em c)} initial rms Alfv\'en {\revised number
$A=v/v_A=5$ MHD} runs with ZEUS, and {\em d)} $A=1$ MHD runs with ZEUS
(From~\cite{m98})}
\label{prlres}
\end{figure}

We next performed resolution studies using ZEUS for three different
cases with no field, weak field and strong field as shown in
Fig.~\ref{prlres}.
The weak field models have an initial ratio of thermal to magnetic
pressure $\beta = 2$, while the strong field models have $\beta =
0.08$.  We ran the same unmagnetized model with the SPH code to
demonstrate that our results are truly independent of the details of
the viscous dissipation, and so that our lack of an explicit viscosity
does not affect our results. 

The kinetic energy decay curves for the four resolution studies are
shown in Fig.~\ref{prlres}.  For each of our runs we performed a
least-squares fit to the slope of the power-law portion of the kinetic
energy decay curves.  These results appear converged at the 5--10\%
level; it is very reassuring that the different numerical methods
converge to the same result for the unmagnetized case.  

We find that highly compressible, isothermal turbulence decays close
to linearly in time, with $\eta = 0.98$.  Adding magnetic fields
decreases the decay rate only slightly to $\eta \sim $0.85--0.9,
with very slight dependence on the field strength or adiabatic index.
{\revised Similar results have been reported by Stone, Ostriker, \&
Gammie~\cite{sog98} and Padoan \& Nordlund~\cite{pn99} for
compressible MHD turbulence, and by Biskamp \& M\"uller~\cite{bm00} for
incompressible MHD turbulence.}

The clear astrophysical implication of these models is that even
strong magnetic fields, with the field in equipartition with the
kinetic energy, cannot prevent the decay of turbulent motions on
dynamical timescales.  {\revised If molecular clouds live for longer
than their dynamical time of roughly a megayear, as is generally
believed even by those arguing for lifetimes under
10~Myr~\cite{bhv99,hbb01} rather than the more classical
30~Myr~\cite{bs80}, then
the significant kinetic energy observed in their}
gas must be supplied more or less continuously.

\section{Driven Turbulence}

\begin{figure}[thbf]
\begin{center}
\includegraphics[width=.8\textwidth]{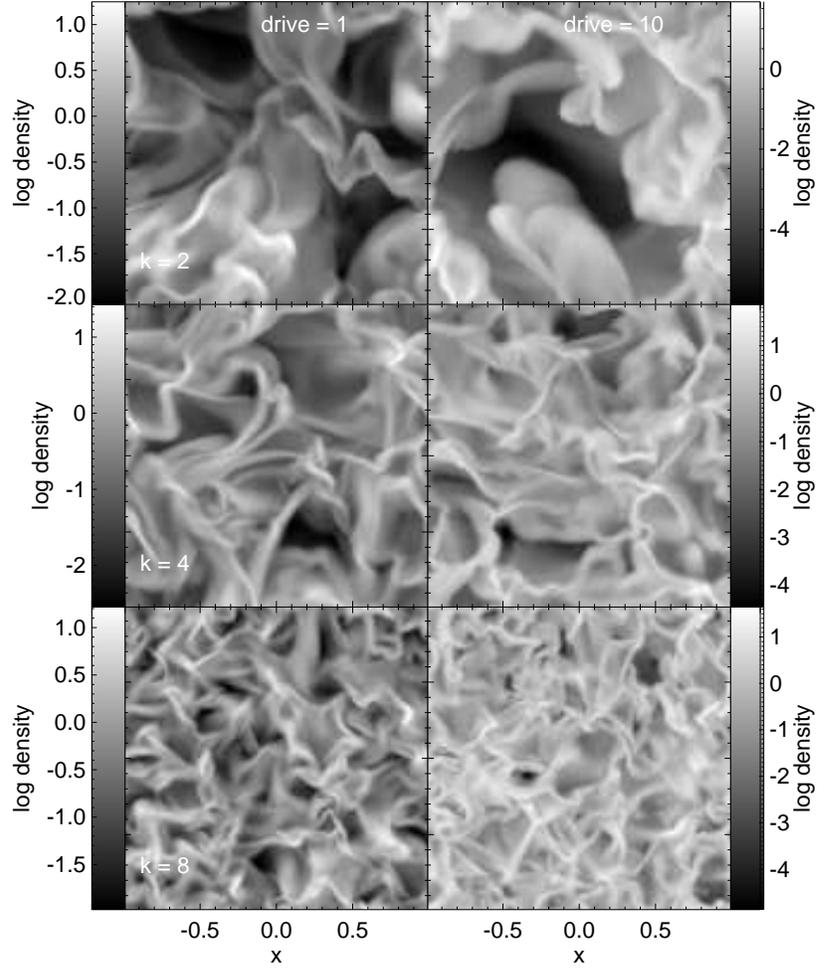}
\end{center}
\caption[Driven isothermal turbulence morphology]{Models showing the
appearance of turbulence with different characteristic size scales and
driving energy input.  Log of density on slices through hydrodynamic
models from~\cite{m99} at a resolution of $128^3$ grid points.  The
value of ``drive'' given in the figure is the driving energy input
rate $\dot{E}_{\rm in}$ for that model, while $k = L/\lambda_d$ is the
wavenumber of the driving pattern, and the size of the cube $L = 2$
for all runs}
\label{drivensix}
\end{figure}

To compute the energy dissipation from uniformly driven turbulence we
initialize the turbulent flow with a narrow band of $k$ values, using
a top-hat function with roughly the same behavior as the steeply
peaked curve used by Stone et al. \cite{sog98} in most of their
models.  We set the velocity field up as described
above~\cite{m98}. To drive the turbulence, we then normalize this
fixed pattern to produce a set of perturbations
$\delta\vec{\nu}(x,y,z)$, and at every time step add a velocity field
$\delta\vec{v}(x,y,z) = A \delta\vec{\nu}$ to the velocity $\vec{v}$,
with the amplitude $A$ now chosen to maintain constant kinetic energy
input rate $\dot{E}_{\rm in} = \Delta E / \Delta t$.  For compressible
flow with a time-dependent density distribution, maintaining a
constant energy input rate requires solving a quadratic equation in
the amplitude $A$ at each time step. For a grid with $N$ zones on a
side, each of volume $\Delta V$, the equation for $A$ is
\begin{equation}
\Delta E = \frac12 \Delta V \sum_{i,j,k = 1}^{N} \rho_{ijk} A \delta
\vec{\nu}_{ijk} \cdot (\vec{v}_{ijk} + A \delta\vec{\nu}_{ijk}).
\end{equation}
We take the larger root of this equation to get the value of $A$.  The
resulting flow is shown in Fig.~\ref{drivensix}.

\begin{figure}[thbf]
\begin{center}
\includegraphics[width=.8\textwidth]{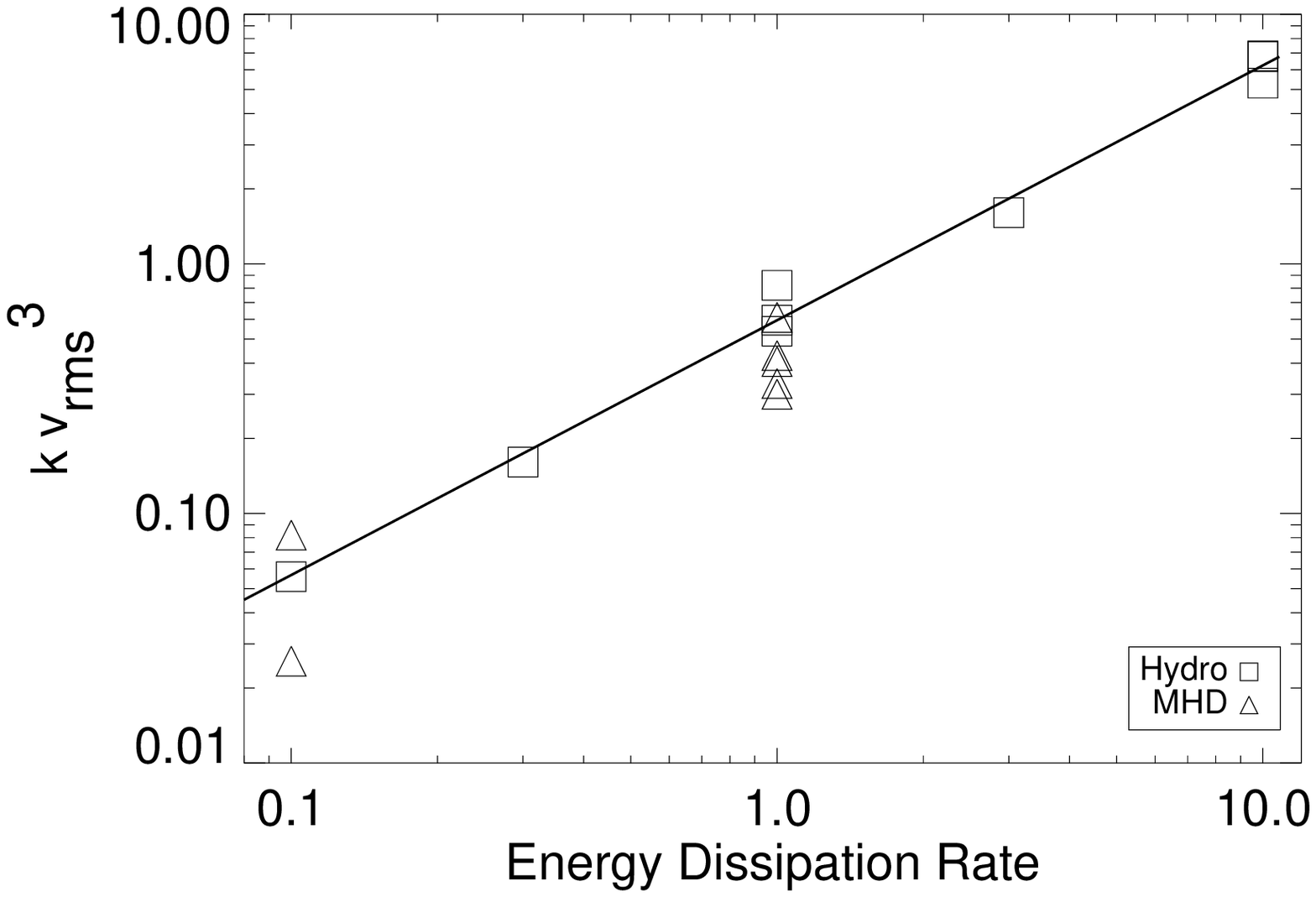}
\end{center}
\caption[Energy dissipation rates]{Energy dissipation rate for hydro
and MHD models compared to $k_d v_{\rm rms}^3$ where $k_d$ is the
dimensionless {\revised driving} wavenumber, and the size of the cube
$L = 2$ for all runs.  The lines have slope of unity, and are fit to
the hydro models, yielding the value for the dissipation coefficient
$\eta_v = 0.21/\pi$ (see equation~[\ref{vdiss}]).  Hydrodynamical
models are indicated by squares, MHD models by triangles
(From~\cite{m99})}
\label{ecalfig}
\end{figure}

We find that the best description of these compressible models comes
by taking a length scale ${\cal L} = \lambda_d$ the driving
wavelength, and a velocity scale ${\cal V} = v_{\rm rms}$ the rms
velocity, rather than any of the other options available. {\revised
Happily, these are just the length and velocity scales that would be
expected from the theory of {\em incompressible} turbulence.}
Figure~\ref{ecalfig} shows
equilibrium energy dissipation rates for all the models described by
Mac Low~\cite{m99}, compared to the quantity $k_d v_{\rm rms}^3 \sim
v_{\rm rms}^3/\lambda_d$.  A fit to the hydrodynamic models
gives a relation with slope 1.02.  Let us define a
dimensionalized wavenumber $\tilde{k} = (2\pi/L) k_d = 2\pi/\lambda_d$.
A very good approximation is then the linear relation
\begin{equation} \label{vdiss}
\dot{E}_{\rm kin} \simeq -\eta_v m \tilde{k} v_{\rm rms}^3,
\end{equation}
with $\eta_v = 0.21/\pi$, where the assumption is made that in
equilibrium $\dot{E}_{\rm kin} = \dot{E}_{\rm in}$.  The dependence on
the mass of the cube $m$ comes strictly from dimensional arguments, as
all of the runs shown have the same mass $m = \rho_0 L^3 =
8$. {\revised The referee of this review points out that $\eta_v k_d
\simeq 0.42 \lambda_d^{-1}$, remarkably close to the value of
$1/(2\lambda_d)$ that would be expected by direct application of the
theory of incompressible turbulence.  This implies that the density
variations average out almost independently of velocity variations.}

\begin{figure}[thbf]
\begin{center}
\includegraphics[width=.8\textwidth]{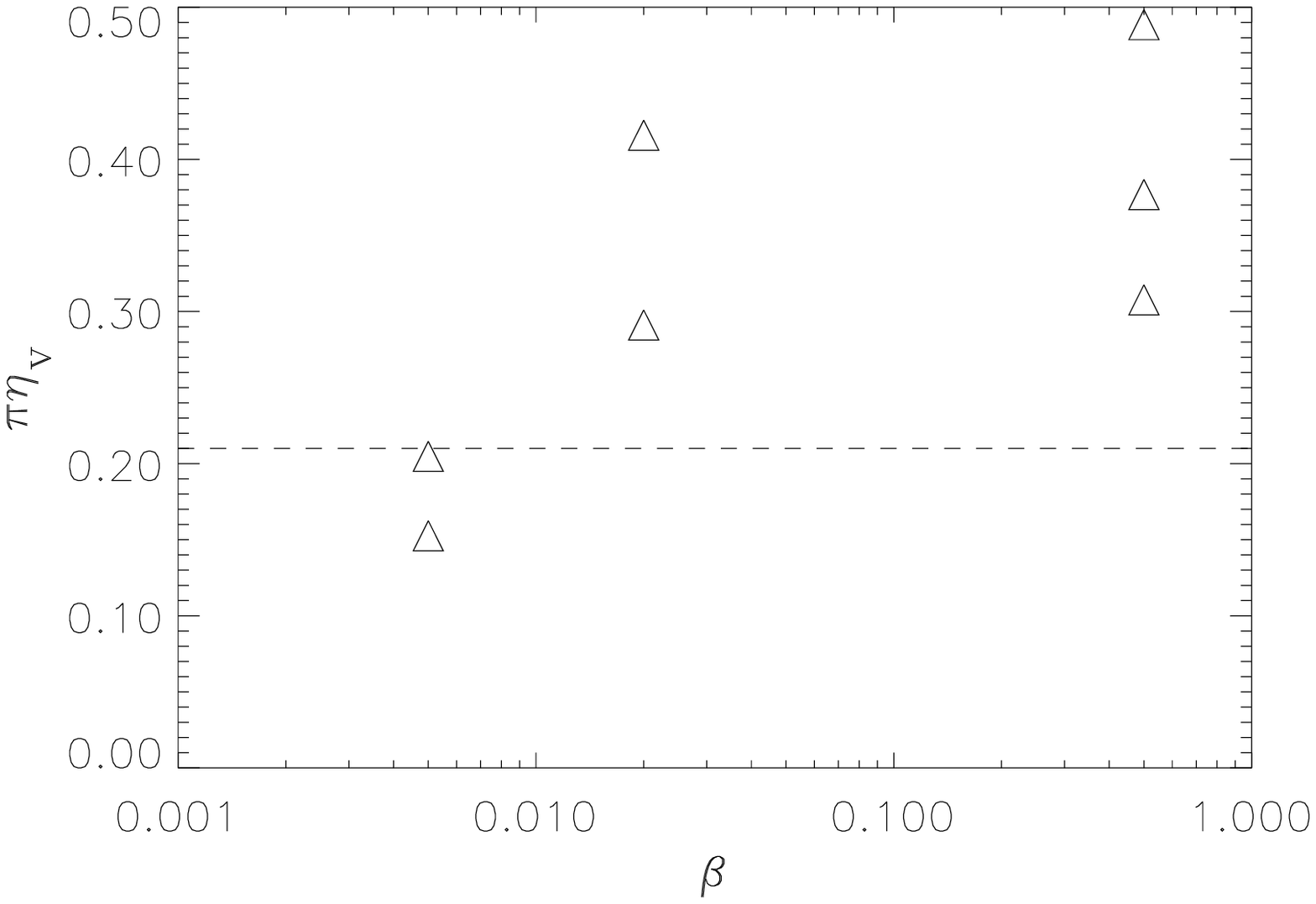}
\end{center}
\caption[Energy dissipation as function of field strength]{Dependence
of the dissipation coefficient $\eta_v = \dot{E}_{\rm kin} / (\tilde{k} m
v_{\rm rms}^3)$ on the plasma $\beta$, the ratio of thermal to
magnetic pressure.  The dashed line shows the value derived from
unmagnetized models.  Models with weaker fields appear to have as much
as a factor two higher dissipation rate, while strong field models
approach the unmagnetized rate (From~\cite{m99})}
\label{ecalfigb}
\end{figure}

The MHD models that fit equations~(\ref{vdiss}) most
closely are the strong field cases, with $v_A/c_s = 10$.  The weak
field cases appear to follow a relation similar to
equation~(\ref{vdiss}), but with values of $\eta_v$ up to a factor of
two higher, as shown in Figure~\ref{ecalfigb}. 
Without further computation, it remains unclear how much of the
variation seen among the models with the same $\beta$ is due to random
fluctuations or the remaining lack of numerical convergence, and how
much is real.  The higher dissipation seen in the high-$\beta$,
weak-field cases could be explained by noting that weak fields will be
more strongly influenced by the flow, generating more dissipative MHD
waves.

We can use these results to discuss whether decaying turbulence
can delay gravitational collapse.  This can formally be
examined by determining whether the ratio
\begin{equation} \tau = t_d / t_{\rm ff} > 1, \label{tau1} \end{equation} 
where the turbulent decay time $t_d = E_{\rm kin} /
\dot{E}_{\rm kin}$, and the free-fall time $t_{\rm ff}$ for the gas is
given by equation~(\ref{equ:tff}).  Because $t_d$ depends not only on the
strength of the turbulence, but also on the driving wavelength, the
value of $\tau$ also depends on the ratio
\begin{equation} \kappa  = \lambda_d / \lambda_J, \label{kappa1} \end{equation}
where the Jeans wavelength $\lambda_J = c_s \sqrt{\pi/G \rho_0}$.
Numerical models that I disucss in the next section show that
turbulence cannot support the gas against collapse at wavelengths
longer than the driving wavelength \cite{lpp90,khm00}, so that $\kappa
\leq 1$ is necessary for the turbulence to fully support against
collapse.

Substituting for the values in equation~(\ref{tau1}), we can write
\begin{equation}
\tau = \frac{E_{\rm kin}}{\dot{E}_{\rm kin}} \frac{c_s}{\lambda_J}
\sqrt{\frac{32}{3}}. 
\end{equation}
We can now use equation~(\ref{vdiss}) for $\dot{E}_{\rm kin}$, and,
somewhat less accurately, take $E_{\rm kin} \sim m v^2_{\rm rms} / 2$,
noting that this introduces no more than a 20--30\% error.
Substituting and using the definition of $\kappa$ given in
equation~(\ref{kappa1}), I find that the dissipation time scaled in
units of the free fall time is
\begin{equation}
\tau(\kappa) = \frac{1}{4 \pi \eta_v} \left(\frac{32}{3}\right)^{1/2}
\frac{\kappa}{M_{\rm rms}} \simeq \,3.9 \,\frac{\kappa}{M_{\rm rms}},
\end{equation}
where $M_{\rm rms} = v_{\rm rms}/c_s$ is the rms Mach number of the
turbulence.  In molecular clouds, $M_{\rm rms}$ is typically observed
to be of order 10 or higher, while $\kappa < 1$ is required for
support, so turbulence will decay long before the cloud collapses and
not markedly influence its collapse.

\section{Self-Gravitating Turbulence}

Now that it is clear that driven turbulence must be present in
molecular clouds, we can ask whether it alone is sufficient to support
clouds against gravitational collapse.

The virial theorem provides a first way of examining this question.
In equilibrium the total kinetic energy in the system adds up to half
its potential energy, $E_{\rm kin} + 1/2\,E_{\rm pot} = 0$. If $E_{\rm
kin} + 1/2\,E_{\rm pot}<0$ the system collapses, while $E_{\rm kin} +
1/2\,E_{\rm pot}>0$ implies expansion. In turbulent clouds, the total
kinetic energy includes not only the internal energy but also the
contributions from turbulent gas motions. If this is taken into
account, simple energy considerations can already provide a
qualitative description of the collapse behavior of turbulent
self-gravitating media \cite{b87}.

A more thorough investigation, however, requires a linear stability
analysis.  For the case of an isothermal, infinite, homogeneous,
self-gravitating medium at rest (i.e.~without turbulent motions)
Jeans~\cite{j02} derived a relation between the oscillation frequency
$\omega$ and the wave number $k$ of small perturbations,
\begin{equation}
\omega^2 - c_{\rm s}^2 k^2 + 4\pi G\,\rho_0 = 0\;,
\label{eqn:jeans-dispersion-rel}
\end{equation}
where $c_{\rm s}$ is the isothermal sound speed, $G$ the gravitational
constant, and $\rho_0$ the initial mass density. Note that the
derivation includes the ad hoc assumption that the linearized version
of the Poisson equation describes only the relation between the
perturbed potential and the perturbed density, neglecting the potential of
the homogeneous solution.  This is the so-called `Jeans
swindle'.  Perturbations are unstable against gravitational contraction
if their wave number is below a critical value, the Jeans wave number
$k_{\rm J}$, i.e.~if
\begin{equation}
k^2 < k_{\rm J}^2 \equiv \frac{4 \pi G \rho_0}{c_{\rm s}^2}\;, 
\label{eqn:jeans-wave-number}
\end{equation}
or equivalently if the wave length of the perturbation exceeds a
critical size given by $\lambda_{\rm J} \equiv 2 \pi k_{\rm J}^{-1}$.
This directly translates into a mass limit. All perturbations with
masses exceeding the Jeans mass,
\begin{equation}
M_{\rm J} \equiv  \rho_0 \lambda^3 =  \left( \frac{\pi}{G}
\right)^{3/2} \rho_0^{-1/2} {c_{\rm s}^3} ,
\label{eqn:jeans-mass}
\end{equation}
will collapse under their own weight.

These and subsequent analytical approaches (reviewed in \cite{khm00})
make a strong assumption that substantially limits their reliability,
namely that the equilibrium state is homogeneous, with constant
density $\rho_0$. However, observations clearly show that molecular
clouds are extremely non-uniform. One way to achieve progress and
circumvent the restrictions of a purely analytical approach is to
perform numerical simulations.
Bonazzola et al.~\cite{b87}, for example, used low resolution ($32
\times 32$ collocation points) calculations with a 2-dimensional
spectral code to support their analytical results.  Also restricted to
two dimensions were the hydrodynamical studies by Passot et
al.~\cite{ppw88}, L{\'e}orat, Passot \& Pouquet~\cite{lpp90},
V{\'a}zquez-Semadeni et al.~\cite{vpp95} and Ballesteros-Paredes,
V{\'a}zquez-Semadeni \& Scalo~\cite{bvs99}, although performed with
far higher resolution.  Magnetic fields were introduced in two
dimensions by Passot, V\'azquez-Semadeni, \& Pouquet~\cite{pvp95}, and
extended to three dimensions with self-gravity (though at only $64^3$
resolution) by V\'azquez-Semadeni, Passot, \& Pouquet~\cite{vpp96}.  A
careful analysis of 1-dimensional computations including both MHD and
self-gravity was presented by Gammie \& Ostriker~\cite{go96}, who extended
their work to 2.5 dimensions more recently~\cite{ogs99}.

We use ZEUS-3D and SPH to examine the gravitational stability of
three-dimensional hydrodynamical turbulence at higher resolution than
before, and include magnetic fields using ZEUS-3D.  The use of both
Lagrangian and Eulerian numerical methods to solve the equations of
self-gravitating hydrodynamics in three dimensions (3D) allows us to
attempt to bracket reality by taking advantage of the strengths of
each approach.  This also gives us some protection against
interpreting numerical artifacts as physical effects.

\begin{figure}[fthb]
\begin{center}
\includegraphics[width=0.8\textwidth]{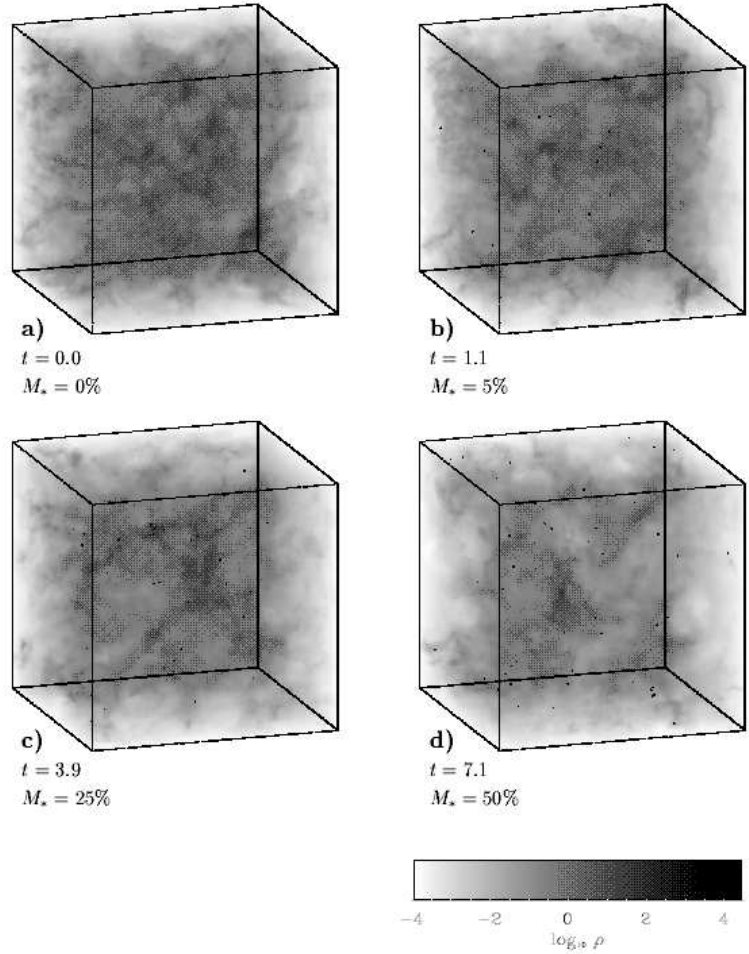}
\end{center}
\caption[Self-gravitating hydro turbulence]{SPH density cubes for
model driven in the wavenumber interval $3\le k_d \le 4$, shown ({\bf
a}) at the time when gravity is turned on, ({\bf b}) when the first
dense cores are formed and have accreted $M_* = 5$\% of the mass,
({\bf c}) when the mass in dense cores is $M_* = 25$\%, and ({\bf d})
when $M_* = 50$\%. Time is measured in units of the global system
free-fall time scale $\tau_{\rm ff}$ (From~\cite{khm00})}
\label{khm-fig1}
\end{figure}

SPH can resolve very high density contrasts because it increases the
particle concentration, and thus the effective spatial resolution, in
regions of high density, making it well suited for computing collapse
problems as shown in Fig.~\ref{khm-fig1}.
By the same token, though, it resolves low-density regions
poorly. Shock structures tend to be broadened by the averaging kernel
in the absence of adaptive techniques.  The correct numerical
treatment of gravitational collapse requires the resolution of the
local Jeans mass at every stage of the collapse~\cite{bb97}.  In the
current code, once an object with density beyond the resolution limit
of the code has formed in the center of a collapsing gas clump it is
replaced by a `sink' particle \cite{bbp95}.  Adequately replacing
high-density cores and keeping track of their further evolution in a
consistent way prevents the time step from becoming prohibitively
small. We are thus able to follow the collapse of a large number of
cores until the overall gas reservoir becomes exhausted.

\begin{figure}[thbf]
\begin{center}
\includegraphics[width=0.7\textwidth,angle=270]{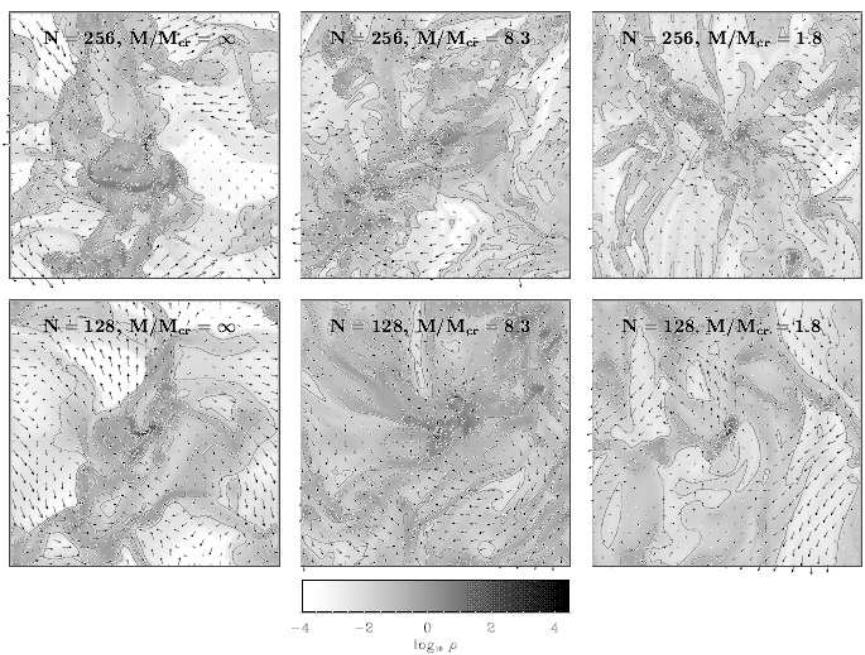}
\end{center}
\caption[Comparison of self-gravitating hydro and MHD
turbulence]{Two-dimensional slices of $256^3$ models from~\cite{hmk01}
driven with wavenumber $k=2$ hard enough that the mass in the box
represents only 1/15 $\langle M_{\rm J}\rangle_{\rm turb}$, and with magnetic
fields strong enough to give critical mass fractions as shown,
corresponding to $\beta = \infty$, 0.9, and 0.04.  Slices are taken at
the location of the zone with the highest density at the time when
$10$\% of the total mass has been accreted onto cores. The plot is
centered on this zone.  Arrows denote velocities in the plane. The
length of the largest arrows corresponds to a velocity of $v \sim 20
c_s$. The density greyscale is given in the colorbar.
(From~\cite{hmk01})}
\label{2dslices}
\end{figure}

ZEUS-3D, conversely, gives us equal resolution in all regions, and
allows us to resolve shocks well everywhere, as well as allowing the
inclusion of magnetic fields, as shown in Fig.~\ref{2dslices}. 
On the other hand, collapsing regions cannot be followed to scales
less than one or two cells.  We must again consider the resolution
required for gravitational collapse.  For a grid-based simulation, the
criterion given by Truelove et al.~\cite{t97} holds. Equivalent to the
SPH resolution criterion, the mass contained in one grid zone has to
be rather smaller than the local Jeans mass throughout the
computation.

Applying this criterion strictly would limit our simulations to the
very first stages of collapse, as we have not implemented anything
like sink particles in ZEUS. We have therefore extended our models
beyond the point of full resolution of the collapse, as we are
primarily interested in the formation of collapsed regions, but not
their subsequent evolution.  Thus, in the ZEUS models, the fixed
spatial resolution of the grid implies that strongly collapsed cores
have a larger cross-section than appropriate for their mass. In
encounters with shock fronts the probability for these cores to get
destroyed or lose material is overestimated.  Cores simulated with
ZEUS are therefore more easily disrupted than they would be
physically. SPH, on the other hand, underestimates the disruption
probability, because sink particles cannot lose mass or dissolve again
once they have formed. The physical result is thus {\em bracketed} by
these two numerical methods.

We use the same driving method as described above for both the SPH and
ZEUS models. We define an effective turbulent Jeans mass $\langle
M_{\rm J}\rangle_{\rm turb}$ by substituting $c_{\rm s}^2
\longrightarrow c_{\rm s}^2 + 1/3 \, \langle v^2 \rangle$ for the
thermal sound speed $c_{\rm s}$ in equation~(\ref{eqn:jeans-mass})
where we approximate the rms velocity of the flow $\langle v^2
\rangle$ by $2E_{\rm kin}/ M$.  The turbulent Jeans mass $\langle
M_{\rm J}\rangle_{\rm turb}$ must be compared to the total system mass
$M\equiv 1$ in order to determine whether global stability is reached.

\begin{figure}[thbf]
\begin{center}
\includegraphics[width=1.0\textwidth]{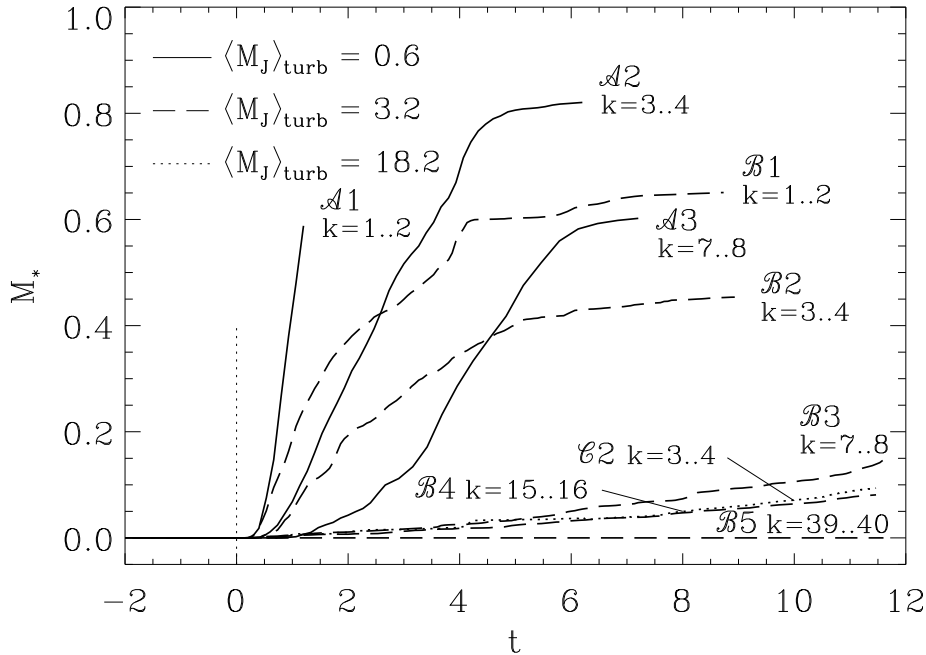}
\end{center}
\caption[Accretion history of driven hydro turbulence]{Fraction of
mass $M_*$ in dense cores as function of time.  All models are
computed using SPH with sink particles replacing dense, collapsed
cores. The different model parameters from~\cite{khm00} are indicated
in the figure. The mass in the box is initially unity, so the solid
curves are unsupported formally, while the others are supported.  The
figure shows how the efficiency of local collapse depends on the scale
and strength of turbulent driving.  Time is measured in units of the
global system free-fall time scale $\tau_{\rm ff}${\revised , while
the driving wavelength $\lambda_d = 2\pi/k$}
(From~\cite{khm00})}
\label{khm-fig4}
\end{figure}

We find that {\em local} collapse occurs even when the turbulent
velocity field carries enough energy to counterbalance gravitational
contraction on global scales, as shown in Fig.~\ref{khm-fig4}.  This
confirms the results of two-dimensional (2D) and low-resolution
($64^3$) 3D computations with and without magnetic fields by
V\'azquez-Semadeni et al.~\cite{vpp96}.  An example of local collapse
in a globally supported cloud is given in Figure~\ref{khm-fig1}.  The
presence of shocks in supersonic turbulence drastically alters the
result from analytic models of incompressible turbulence, as was first
noted by Elmegreen~\cite{bge93} and studied numerically by
V\'azquez-Semadeni et al.~\cite{vpp96}.  The density contrast in
isothermal shocks scales quadratically with the Mach number, so the
shocks driven by supersonic turbulence create density enhancements
with $\delta \rho \propto {\cal M}^2$, where ${\cal M}$ is the {\em
rms} Mach number of the flow. In such fluctuations the local Jeans
mass is {\em decreased} by a factor of $\cal M$ and therefore the
likelihood for gravitational collapse {\em increased}.

To test this explanation numerically, we designed a test case driven
at short enough wave length and high enough power to support even
fluctuations with $\delta \rho \propto {\cal M}^2$, and ran it with
both codes, driven with a wave number $k=39-40$. Within 20 $\tau_{\rm
ff}$ this model shows no signs of collapse.  All the other globally
supported models with less extreme parameters that we computed did
form dense cores during the course of their evolution, supporting our
hypothesis that local collapse is caused by the density fluctuations
resulting from supersonic turbulence.

Magnetic fields might alter the dynamical state of a molecular cloud
sufficiently to prevent gravitationally unstable regions from
collapsing~\cite{mck99}.  They have been hypothesized to support
molecular clouds either magnetostatically or dynamically through MHD
waves.  Mouschovias \& Spitzer~\cite{ms76} derived an expression for the
critical mass-to-flux ratio in the center of a cloud for magnetostatic
support. Assuming ideal MHD, a self-gravitating cloud of mass $M$
permeated by a uniform flux $\Phi$ is stable if the mass-to-flux ratio
\begin{equation}
  \frac{M}{\Phi} <
  \left(\frac{M}{\Phi}\right)_{cr}\equiv\frac{c_\Phi}{\sqrt{G}}.
  \label{equ:magstatsup}
\end{equation}
with $c_\Phi$ depending on the geometry and the field and density
distribution of the cloud.  A cloud is termed {\em subcritical} if it
is magnetostatically stable and {\em supercritical} if it is not.
Mouschovias \& Spitzer~\cite{ms76} determined that $c_\Phi = 0.13$ for
a spherical cloud.

We include magnetic fields in our models of driven, self-gravitating
turbulence to test their effectiveness in supporting against
self-gravity. The MHD simulations start with a uniform magnetic field
in the $z$-direction. We must consider the resolution required to
accurately follow magnetized collapse. Numerical diffusion can reduce
the support provided by a static or dynamic magnetic field against
gravitational collapse.  Increasing the numerical resolution decreases
the scale at which numerical diffusion acts.  For strong, subcritical
fields, the resolution should ensure that numerical diffusion remains
unimportant even for the dense, shocked regions.  We did a suite of
models varying the sound speed and the mass in the cube while holding
the magnetic field strength constant, thus varying $M/M_{cr}$ and the
number of zones in a Jeans length $\lambda_J$. From these models we
conclude that for a self-gravitating magnetostatic sheet to be well
resolved with the algorithm under study, its Jeans length must exceed
four zones~\cite{hmk01}.

\begin{figure}[thbf]
\begin{center}
\includegraphics[width=1.0\textwidth]{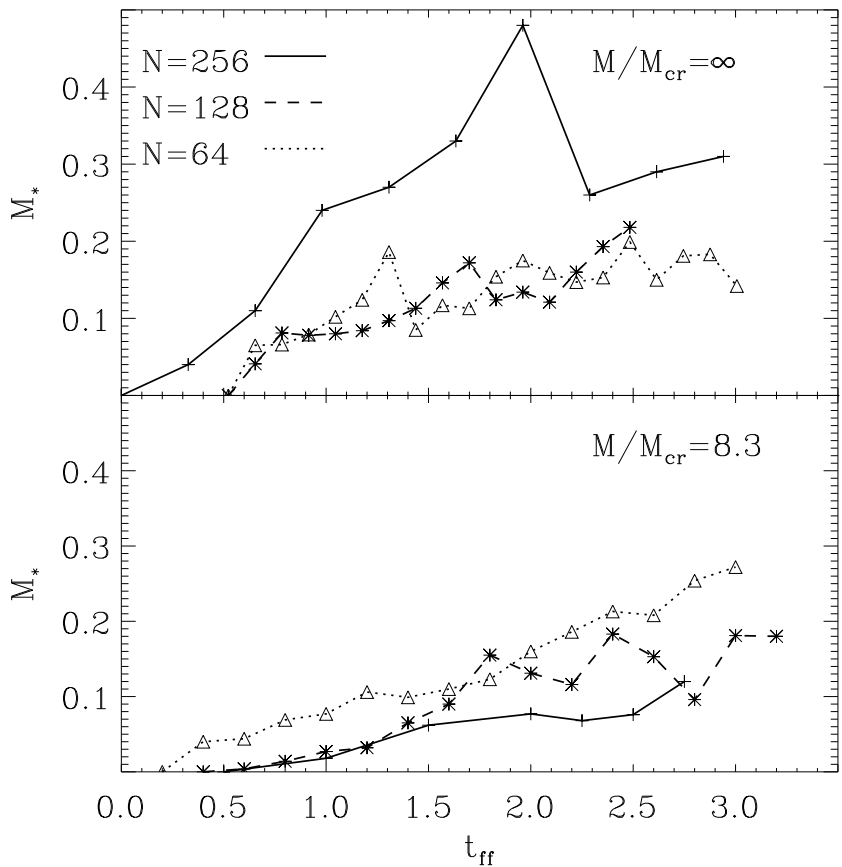}
\end{center}
\vspace{-0.2in}
\caption[Resolution study of self-gravitating hydro and MHD
turbulence]{Comparison of the mass accretion behaviour for runs driven
at wavenumber $k=1-2$ with varying resolution. Pure hydro runs are
shown in the upper panel, and MHD runs in the lower panel, with
resolutions $64^3$ ({\em dotted}), $128^3$ ({\em dashed}), and $256^3$
({\em solid}).  $M_*$ denotes the sum of masses found in all cores
determined by the modified {\tt clumpfind}
algorithm~\cite{wgb94,khm00}. {\revised Note that these cores are
subject to destruction as they can only collapse to the grid scale.}
Times are given in units of free-fall time.  Although the collapse
rate varies, we get collapse in all cases (From~\cite{hmk01})}
\label{massfrac}
\end{figure}

Collapse occurs in both unmagnetized and magnetized cases at all
resolutions, as shown in Figure~\ref{massfrac}.
However, increasing the resolution makes itself felt in different ways
in hydrodynamical and MHD models.  In the hydrodynamical case, higher
resolution results in thinner shocks and thus higher peak densities.
These higher density peaks form cores with deeper potential wells that
accrete more mass and are more stable against disruption.  If we
increase the resolution in the MHD models, on the other hand, we can
better follow short wavelength MHD waves, which appear to be able to
delay collapse, although not to prevent it.  This result extends to
models with $512^3$ zones~\cite{h01}.

\section{Characterization of Turbulence}

\subsection{Wavelet Transforms}

To obtain clues to the true physical nature of interstellar
turbulence, characteristic scales and any inherent scaling laws have
to be measured and modelled. A major problem with characterizing both
the observations and the models is to determine what scaling behaviour,
if any, is present in complex turbulent structures.  Both the velocity
and density fields need to be considered, but only the radial velocity and
column densities can be observed.

\begin{figure}[thbf]
\begin{center}
\includegraphics[width=0.8\textwidth]{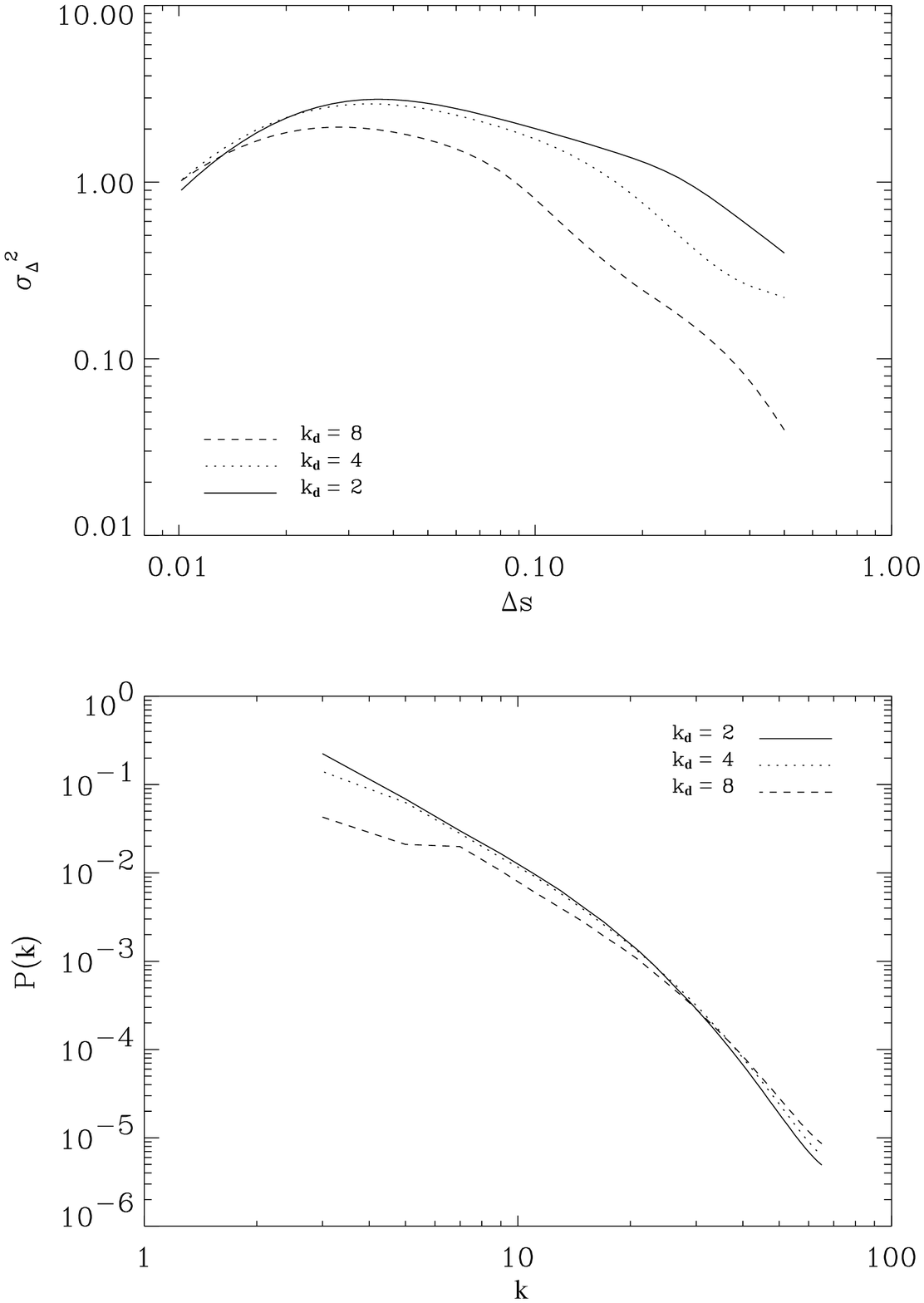}
\end{center}
\caption[Fourier transform vs.\ $\Delta$-variance]{Comparison of 3D
$\Delta$-variance spectra ({\em above}) to Fourier power spectra ({\em
below}) for $128^3$ models of turbulence driven at wavenumbers of $k_d
= 2$ (solid), $k_d = 4$ (dotted), and $k_d = 8$ (dashed),
demonstrating that the analytic relation between power spectrum slope
and $\Delta$-variance spectrum slope also holds for the local behavior
of structures not showing a straight power-law (From~\cite{mo00})}
\label{mo-fig1}
\end{figure}

One measure useful for characterizing structure and scaling in
observed maps of molecular clouds is the $\Delta$-variance,
$\sigma^2_{\Delta}$, an averaged wavelet transform method {\revised
for measuring the amount of structure at different scales} introduced
by Stutzki et al.~\cite{s98}. The $\Delta$-variance spectrum clearly
shows characteristic scales and scaling relations, and its logarithmic
slope can be analytically related to the spectral index of the
corresponding Fourier power spectrum (see Fig.~\ref{mo-fig1}).  In
order to understand the physical significance of the characterization
of the observational maps by $\Delta$-variance spectra, we apply the
same analysis to observations of the  {\revised Polaris Flare at three
different scales~\cite{f98,bp01,ht90}, as analyzed by}~\cite{s98,bso01}
and to simulated {\revised observations of} gas distributions
resulting from our MHD models~\cite{mo00,om02}{\revised, as well as to
the actual density distributions from those models}.

In Fig.~\ref{mo-fig1} we compare the power spectrum and
$\Delta$-variance for {\revised the density distibution from} three
simulations with different driving scales. Each shows a different
characteristic scale formed by the driving, visible as a turn-over at
large lags in the $\Delta$-variances and at small wavenumbers in the
power spectrum, respectively. At smaller lags and higher wavenumbers
power laws can be seen in both cases {\revised for the models with
longer wavelength driving}, with their slopes related by the analytic
relation mentioned above.  A steep drop-off follows at the smallest
scales indicating the resolution limit of the simulation.  {\revised
The model with the shortest wavelength ($k_d = 8$) driving shows no
power law because the driving scale of 16 zones is only marginally
larger than the dissipation scale of about 10 zones.}

The $\Delta$-variance analysis of astronomical maps was extensively
discussed by Bensch et al.~\cite{bso01}. In all the observations
they analyzed, the total cloud size was the only characteristic
scale detected by means of the $\Delta$-variance.  Below that size
they found a self-similar scaling behaviour reflected by a power law
with index $\alpha=$0.5--1.3 corresponding to a Fourier power spectral
index $\zeta=$2.5--3.3.

\begin{figure}[thbf]
\begin{center}
\includegraphics[width=0.9\textwidth]{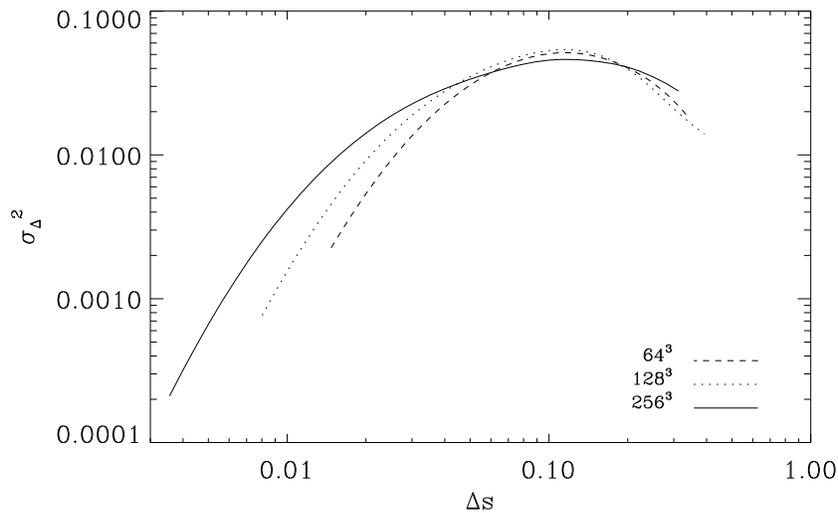}
\end{center}
\caption[Resolution study of $\Delta$-variance]{$\Delta$-variance of
the density cubes (projected to two-dimensions) for a model of
decaying hydrodynamic turbulence at a time $t/t_s = 0.1$, where $t_s$
is the sound-crossing time, computed using numerical resolutions of
$64^3$, $128^3$, and $256^3$ zones (From~\cite{mo00})}
\label{mo-fig4}
\end{figure}

In Fig.~\ref{mo-fig4} we show how numerical resolution, or
equivalently the {\revised dissipation} scale, influences the
$\Delta$-variance spectrum that we find from our simulations.  We test
the influence of the numerical resolution on the structure by
comparing a simple hydrodynamic problem of decaying turbulence
computed at resolutions from $64^3$ to $256^3$, with an initial rms
Mach number $M = 5$.
In contrast to the results from \cite{m99} which showed little
dependence of the energy dissipation rate on the numerical resolution,
we find here clear differences in the scaling behaviour of the
turbulent structures.  At small scales we find a very similar decay in
the relative structure variations up to scales of about 10 times the
pixel size (0.03, 0.06, and 0.1 for the resolutions $256^3$, $128^3$,
and $64^3$, respectively) in all three models. This constant length
range starting from the pixel scale clearly identifies this decay as
being due to the numerical viscosity acting at the smallest available
size scales. 

{\revised Similar behaviour can be observed at the largest lags where
structure variations decay for all three simulations at a length scale
of roughly a quarter the cube size. This structure reflects the
original driving of the turbulence with wavenumbers $1 < k_d < 8$,
which produces structure on the corresponding length scales.
Structures larger than about half the cube size are suppressed by the
periodicity of the simulation cubes.}

\begin{figure}[thbf]
\begin{center}
\includegraphics[width=0.6\textwidth,angle=90]{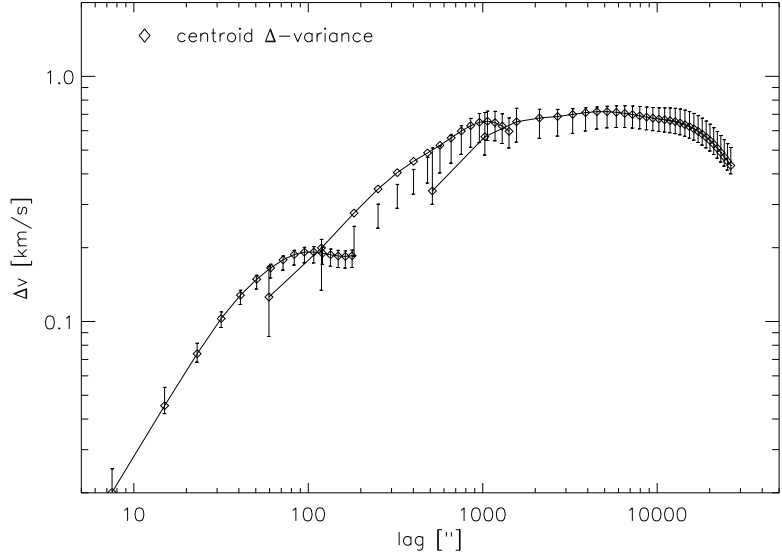}
\end{center}
\caption[Polaris flare $\Delta$-variance spectrum]{The square root of
the $\Delta$-variance of the smoothed Polaris Flare velocity centroid
maps. The size of the smoothing filter was adapted to remove only the
average noise level.  {\revised Spectra are shown from three nested
maps obtained with the 1.2~m CfA telescope with half power beam width
(HPBW) 8.7', the 3~m KOSMA with HPBW 2.2', and the 30~m IRAM with HPBW
0.35'.  At a distance of 150~pc, $1000'' = 0.73$~pc.}  (From~\cite{om02})}
\label{fig-deltavar}
\end{figure}

We can now attempt to interpret the observations using the behavior
seen in the model $\Delta$-variance spectra, and equivalent behavior
seen in velocity centroid $\Delta$-variance spectra of the models,
{\revised as described by Ossenkopf \& Mac
Low}~\cite{om02}. Fig.~\ref{fig-deltavar} shows the square root of the
$\Delta$-variance for the Polaris Flare velocity centroid maps. An
upturn in power at the
smallest lags in each map produced by the observational noise adding
power at small scales was subtracted to prepare these
spectra~\cite{bso01}. The velocity structure in the Polaris Flare maps
shows power-law behaviour from scales of about 0.05 to several
parsecs.  Below that range the slope of the $\Delta$-variance spectrum
of the velocity centroids steepens{\revised, perhaps because of
physical dissipation through ambipolar diffusion or other mechanisms,}
while above that range it flattens {\revised because of the lack of
emission on larger scales}. Bensch et~al.~\cite{bso01} demonstrated
similar scaling behaviour in the intensity maps, reflecting the
density structure.

Comparison between the observations of the Polaris Flare and the
turbulence simulations constrains the mechanisms driving the
turbulence in this cloud. Any mechanism that drives at an intermediate
length scale, such as jets from embedded protostars, should produce
{\revised a peak} in the $\Delta$-variance at that scale, which
is not observed. The approximately self-similar, power-law behaviour
seen in the observations is best reproduced by models where the energy
is injected at large scales and dissipated at small scales.
Driving by interactions with superbubbles and field supernova
remnants~\cite{nf96,m02} would provide such a driving mechanism.  The
Polaris Flare molecular cloud lies in the wall of a large cylindrical
structure representing one of the nearest H~{\sc i} supershells, the
North Celestial Polar Loop~\cite{m91}, adding additional support to
this proposal.

The dominant physical mechanism for dissipation in molecular clouds
was first shown by~Zweibel \& Josafatsson~\cite{zj83} to be ambipolar
diffusion. Klessen et al.~\cite{khm00} showed that the length scale on
which ambipolar diffusion will become important can be found by
examining the ambipolar diffusion Reynolds number
\begin{equation}
R_A = {\cal M}_A \tilde{L} \nu_{ni}/v_A
\end{equation}
defined by Balsara~\cite{b96} and Zweibel \& Brandenburg~\cite{zb97},
where $\tilde{L}$ and ${\cal M}_A $ are the characteristic length and
Alfv\'en Mach number, $\nu_{ni} = \gamma \rho_i$ is the rate at which
each neutral is hit by ions, and $v_A^2 = B^2/4\pi\rho$ approximates
the effective Alfv\'en speed in a mostly neutral region with total
mass density $\rho = \rho_i + \rho_n$ and magnetic field strength $B$.
The coupling constant depends on the cross-section for ion-neutral
interaction, and for typical molecular cloud conditions has a value of
$\gamma \approx 9.2 \times 10^{13}$~cm$^3$~s$^{-1}$~g$^{-1}$
(e.g.~\cite{sm97}). 

Setting the ambipolar diffusion Reynolds number $R_A = 1$ yields a
diffusion length scale of
\begin{eqnarray}
L_D & = & v_A / {\cal M}_A \nu_{ni} \\
    & \approx & (0.041\mbox{pc}) {\cal M}_A 
      \left(\frac{B}{10\mbox{ $\mu$G}}\right)\!  
      \left(\frac{10^{-6}}{x}\right)\!
      \left(\frac{10^3 \mbox{ cm}^{-3}}{n_n}\right)^{3/2}
\end{eqnarray}
with the ionization fraction $x = \rho_i/\rho_n$ and the neutral
number density $n_n = \rho_n/\mu$, with $\mu = 2.36 m_H$.  If the
ionization level in the Polaris Flare is low enough and the
field is high enough, this length scale of order 0.05 pc
would be directly resolved in the IRAM observations.  We cannot yet
unambiguously say whether the steepening observed at the smallest
scales in the velocity spectra is due to beam smearing at the
observational resolution limit or to a detection of the actual
dissipation scale, similar to the downturn at the dissipation scale in
the numerical models. If better observations do continue to show such
a downturn in the future, that will be an indication of the
dissipation scale.

\subsection{Clump Characterization}

Using 2D numerical simulations, Ballesteros-Paredes et
al.~\cite{bvs99} showed that observed clumps frequently come from the
superposition of several physically disconnected regions in the line
of sight at the same radial velocity (see also \cite{osg01}) but not
necessarily at the same position or three-dimensional velocity.
{\revised This had earlier been suggested by workers including
Burton~\cite{b71}, Issa et al.~\cite{imw90}, Falgarone, Puget, \&
P\'erault~\cite{f92}, and Adler \& Roberts~\cite{ar92}.}

We used the models described above to examine this effect in 3D
using simulated observations.  We calculate the radiative
transfer through the density and velocity fields given by the
computations, assuming local thermodynamic equilibrium (LTE) for the
population of the molecular energy levels, which is a sufficiently
good approximation to study qualitatively the effects of projection
and superposition (velocity crowding) of structure in the cloud. In
order to determine differences when observing the same region with
different tracers, we also set minimum density thresholds below which
the molecules might be underexcited.

Velocity crowding contributes to the generation of clumps in the
observational space, so observed clumps frequently contain emission
from physically separated regions~\cite{bvs99,p00,osg01,l01}. We
demonstrate this effect using a typical MHD simulation with
intermediate values of driving strength
($L=1$) and wavenumber ($k=4$). 

\begin{figure}[thbf]
\begin{center}
\includegraphics[width=1\textwidth]{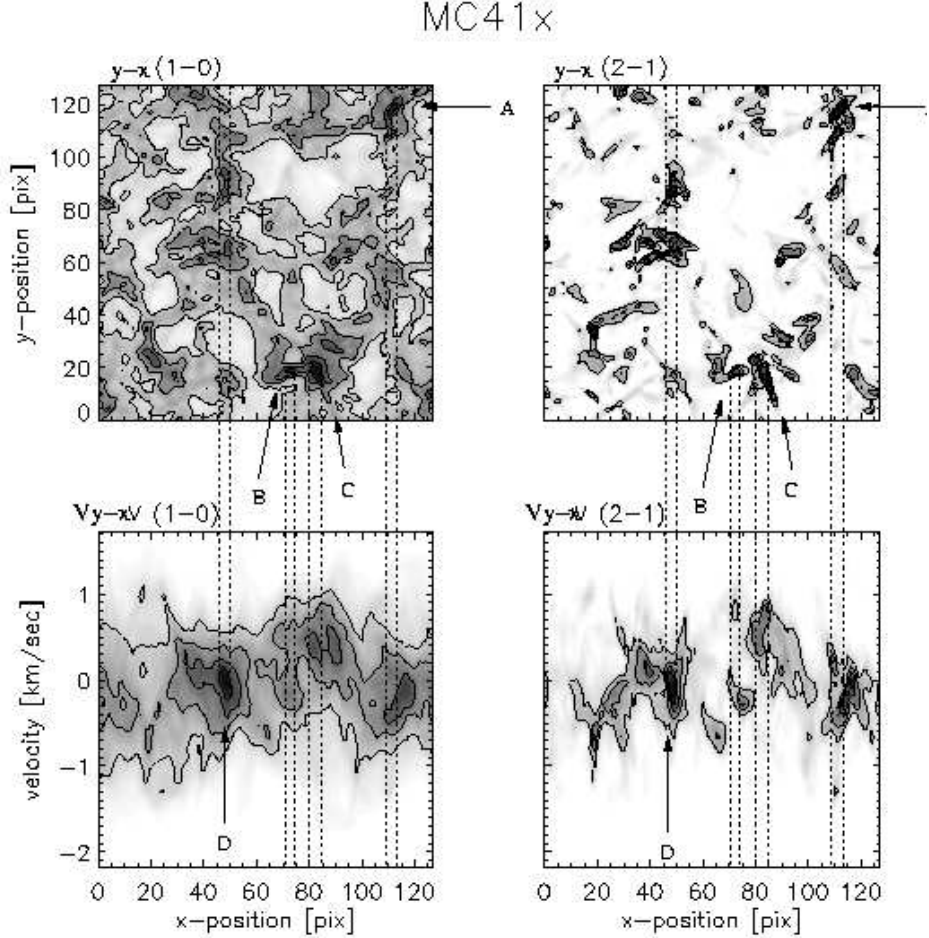}
\end{center}
\caption[Difference between physical and observed clumps]{Physical and
velocity space maps for $^{13}$CO(1-0) and (2-1) in a run with
$\dot{E}=1$, $k=4$ and $B=0.1$. Clumps in physical space (A, B, C) do
not necessarily correspond to clumps in the observed velocity
space. Observed clumps in velocity space (D) are not necessarily
formed by emission from a single region in physical space.
From~\cite{bm02}}
\label{jbp-fig6}
\end{figure}

Figure~\ref{jbp-fig6} shows physical and observed maps for
$^{13}$CO(1-0) and (2-1). We see that clumps in real space (letters A,
B, and C in panel $y-x(1-0)$) do not necessarily have a counterpart in
observational space.  Clumps in observational space, on the other
hand, (letter D in panel $Vy-x(1-0)$) do not necessarily come from isolated
regions in real space, but have contributions from many different
regions along the same line of sight. In Figure~\ref{jbp-fig6} we plot
dotted lines that show the places where the emission of a physical
clump lies in observational space, and where the emission from an
observational clump is generated in physical space. For reference, we
use the same lines in the $^{13}$CO(1-0) map as in the $^{13}$CO(2-1)
map. 

If clumps in the observational space are the result of the
contribution of multiple regions in the physical space it is of
primary importance to understand whether relationships reported for
observed clumps in molecular clouds are also valid for the actual
physical clumps. 

Larson~\cite{l81} studied the dependence with size of the mean density,
velocity dispersion, and mass spectrum of the clouds in a sample of
observational data taken from the literature. He found 
\begin{eqnarray}
\rho & \propto R^\alpha,
\label{rhomean} \\
\delta v & \propto R^m,
\label{deltav} \\
{dN\over d \log{M}} & \propto M^\gamma.
\label{espectro}
\end{eqnarray}
The most commonly quoted values in the literature are $\alpha\sim -1$,
$m\sim 1/2$, and $\gamma\sim -0.5$. However, there is some discrepancy
in the values reported~\cite{c87,l89}.

Larson~\cite{l81} himself mentioned that the relationships he found
might be due to the limitations of the observations. Kegel~\cite{k89}
first demonstrated that the observed mean density-size relationship
could be due to observational effects, and that the observed and
physical {\revised values of properties} such as radius or volumetric
density might be quite different. Scalo~\cite{s90} showed that CO and
extinction saturate at roughly the same column density, which forms
the upper envelope of the mean density-size relationship.
V\'azquez-Semadeni et al.~\cite{vbr97} reported the lack of a mean
density-size relationship, confirming numerically the analysis by
Kegel (\cite{k89}, see also the discussions in \cite{l81} and
\cite{s90}) in the sense that there are clouds with small sizes and
low column density that will be undetected in observational surveys.

\begin{figure}[thbf]
\begin{center}
\includegraphics[width=1\textwidth]{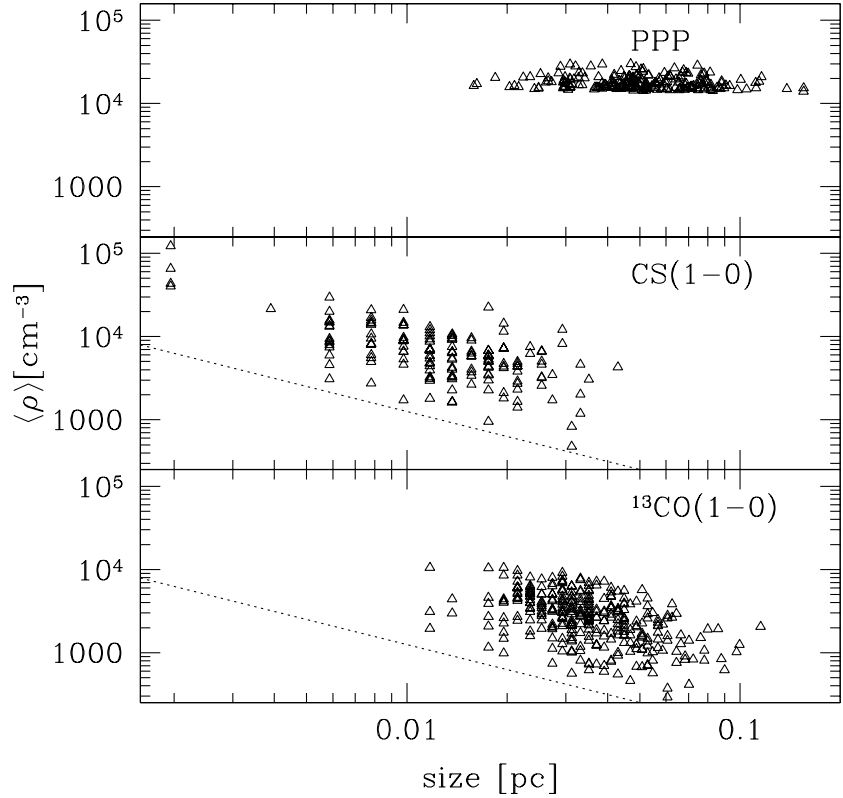}
\caption[Mean density-size relation as observational artifact]{Mean
density-size relationship for physical clumps in physical (PPP, upper
panel); and simulated observational clumps in two different density
tracers in observational coordinates (PPV, middle and lower
panels). The dotted line has a slope of $\alpha = -1$. In physical
space we find no correlation, verifying the results of
V\'azquez-Semadeni et al.~\cite{vbr97}, but nevertheless the simulated
observations show such a correlation, as found by Larson~\cite{l81}
and many others. The selection of two different density tracers was
chosen to show that the apparent correlation does not depend on the
selection of the density threshold. (From \cite{bm02})
\label{jbp-fig9}}
\end{center}
\end{figure}

In Figure~\ref{jbp-fig9} we use the computational model described
above to compare the density-size relationships in physical and
observational spaces.  We define clumps as a connected set of points
below a local maximum following the intensity {\em only} downwards
until the threshold is reached, a scheme implemented in the code
called {\tt clumpfind}~\cite{wgb94}.
We note three points about Figure~\ref{jbp-fig9}. First, there is no
relation between mean density and size (eq.~[\ref{rhomean}]),
confirming the 2D results of V\'azquez-Semadeni et al.~\cite{vbr97}.
Second, there is a minimum density below which there are no clumps
identified.  This minimum is just given by the density threshold we
used in {\tt clumpfind}. Third, even though the simulations exhibit a large
dynamical range in density ($\rho_{\rm max}/\rho_{\rm min} \sim 3.5
\times 10^4$), the dynamical range in the mean density-size
relationship is small, because in constructing such a plot we choose
the clumps around the local density maxima.

In the middle and lower panels of Fig.~\ref{jbp-fig9} we show the mean
density-size relationship for clumps in simulated observational maps
of the same run, integrated along the $x$ axis. The middle panel is the
one obtained by using CS(1-0), and the lower panel is the relationship
obtained by using $^{13}$CO(1-0).  The observed clumps do exhibit
approximately the relationship given by Larson~\cite{l81}, despite the
lack of correlation exhibited by the physical clumps in this model.
We conclude from this demonstration that the observed density-size
relationship (eq.~[\ref{rhomean}]) is an observational artifact.

{\revised Kegel~\cite{k89} suggested one mechanism that appears able
to explain our results: only clouds with emission exceeding some
intensity threshold will be detected, effectively setting a column
density cutoff, rather than the physical density cutoff imposed in the
physical density-size relationship. (This threshold is determined by
the noise level in observations, or by some average intensity in the
simulated observations). A constant column density cutoff produces a
cutoff with slope $-1$ in the mean density-size plane (middle and
lower panels), just as the constant physical density cutoff in
physical space produces a flat cutoff.}

\section{Supernova-Driven Turbulence}

Both support against gravity and maintenance of observed motions
appear to depend on continued driving of the turbulence, as I have
described.  What then is the energy source for this driving?

Motions coming from gravitational collapse have often been suggested,
but fail due to the quick decay of the turbulence as described above.
If the turbulence decays in less than a free-fall time, then it cannot
delay collapse for substantially longer than a free-fall
time~\cite{kb00}. 

Protostellar jets and outflows are another popular suspect for the
energy source of the observed turbulence.  They are indeed quite
energetic, but they deposit most of their energy into low density gas,
as is shown by the observation of multi-parsec long jets extending
completely out of molecular clouds~\cite{bd94}.  Furthermore, as
described in the previous section, the observed motions show
increasing power on scales all the way up to and perhaps beyond the
largest scale of molecular cloud complexes~\cite{om02}.  It is hard to
see how such large scales could be driven by protostars embedded in
the clouds.

Another energy source that has long been considered is shear from
galactic rotation.  Work by Sellwood \& Balbus~\cite{sb99} has shown
that magnetorotational instabilities~\cite{bh91,bh98}
could couple the large-scale motions to small scales efficiently.  For
parameters appropriate to the far outer H~{\sc i} disk of the Milky
Way, they derive a resulting velocity dispersion of 6~km~s$^{-1}$,
close to that observed.  This instability may provide a base value for the
velocity dispersion below which no galaxy will fall.  If that is
sufficient to prevent collapse, little or no star formation will
occur, producing something like a low surface brightness galaxy with
large amounts of H~{\sc i} and few stars.

In active star-forming galaxies, however, clustered and field
supernova explosions appear likely to dominate the driving, raising
the velocity dispersion to the 10--15 km~s$^{-1}$ observed in
star-forming portions of galaxies (see work cited in \cite{m00} for
example). {\revised These explosions will be predominantly from
B~stars no longer associated with their parent gas, as they are far
more numerous than more massive O~stars, have explosions that are just
as powerful, and live long enough (up to 50~Myr) to either drift away
from or ionize their parent clouds.}  This provides a large-scale
self-regulation mechanism for star formation in disks with sufficient
gas density to collapse despite the velocity dispersion produced by
the magnetorotational instability.  As star formation increases in
such galaxies, the number of OB stars increases, ultimately increasing
the supernova rate and thus the velocity dispersion, which will
restrain further star formation.

Supernova driving not only determines the velocity dispersion, but may
actually form molecular clouds by sweeping gas up in a turbulent
flow. Clouds that are turbulently supported will experience
inefficient, low-rate star formation, while clouds that are too
massive to be supported will collapse~\cite{ko01}, undergoing
efficient star formation to form OB associations or even starburst
knots.

We study supernova driving in a magnetized medium numerically, using
the RIEMANN framework for computational astrophysics, which is based
on higher-order Godunov schemes for MHD \cite{rb96,b98a,b98b}, and
incorporates schemes for pressure positivity \cite{bs99a}, and
divergence-free magnetic fields \cite{bs99b}.  (The framework also
includes parallelized, divergence-conserving, MHD adaptive mesh
refinement \cite{bn01,b01}, though no results using that capability
are shown here.) In the models presented here, we solve the ideal MHD
equations including both radiative cooling and pervasive heating in a
(200 pc)$^3$ periodic computational box, using a grid of 128$^3$
cells. We start the simulations with a uniform density of $2.3 \times
10^{-24}$ g cm$^{-3}$, threaded by a uniform magnetic field in the
$x$-direction with strength 5.8 $\mu$G, a factor of roughly two
stronger than that observed in the Milky Way disk.  This very strong
field maximizes the effects of magnetization on the turbulence.

For the cooling, we use a tabulated version of the radiative cooling
curve shown in Figure~1 of MacDonald and Bailey~\cite{mb81}, which is
based on the work of Raymond, Cox \& Smith~\cite{rcs76} and Shapiro and
Moore~\cite{sm76}.  (It falls smoothly from temperatures of order $10^5$~K
to $10^2$~K, not incorporating a sharp cutoff at $10^4$~K due to the
turnoff of Ly$\alpha$ cooling.)  In order to prevent the gas from
cooling below zero, we set the lower temperature cutoff for the
cooling at 100~K.  We also include a diffuse heating term to represent
processes such as photoelectric heating by starlight, which we set
constant in both space and time.  We set the heating level such that
the initial equilibrium temperature determined by heating and cooling
balance is 3000~K.

\begin{figure}[thbf]
\centerline{\hbox{
\includegraphics[width=0.5\textwidth,angle=270]{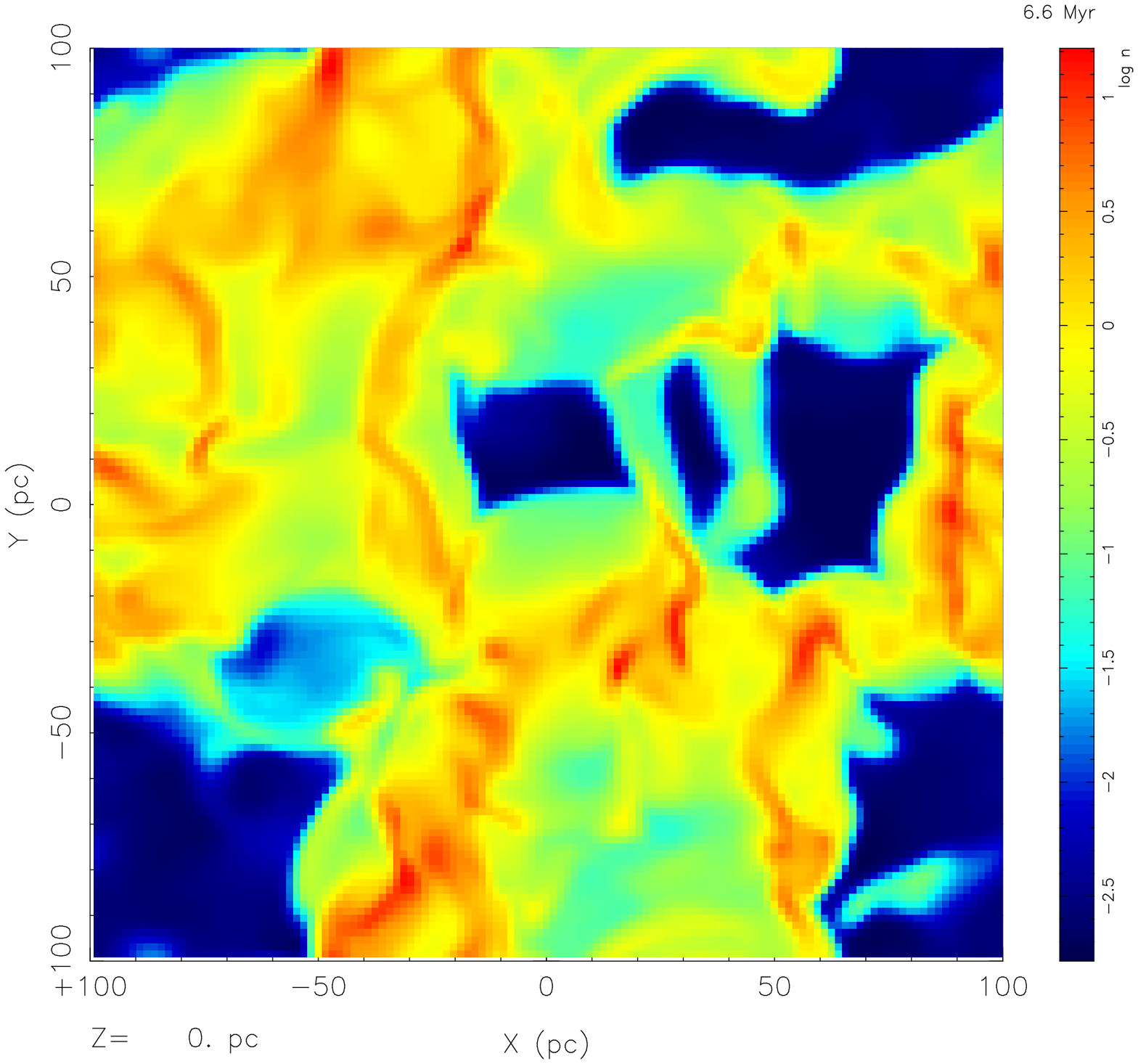}
\includegraphics[width=0.5\textwidth,angle=270]{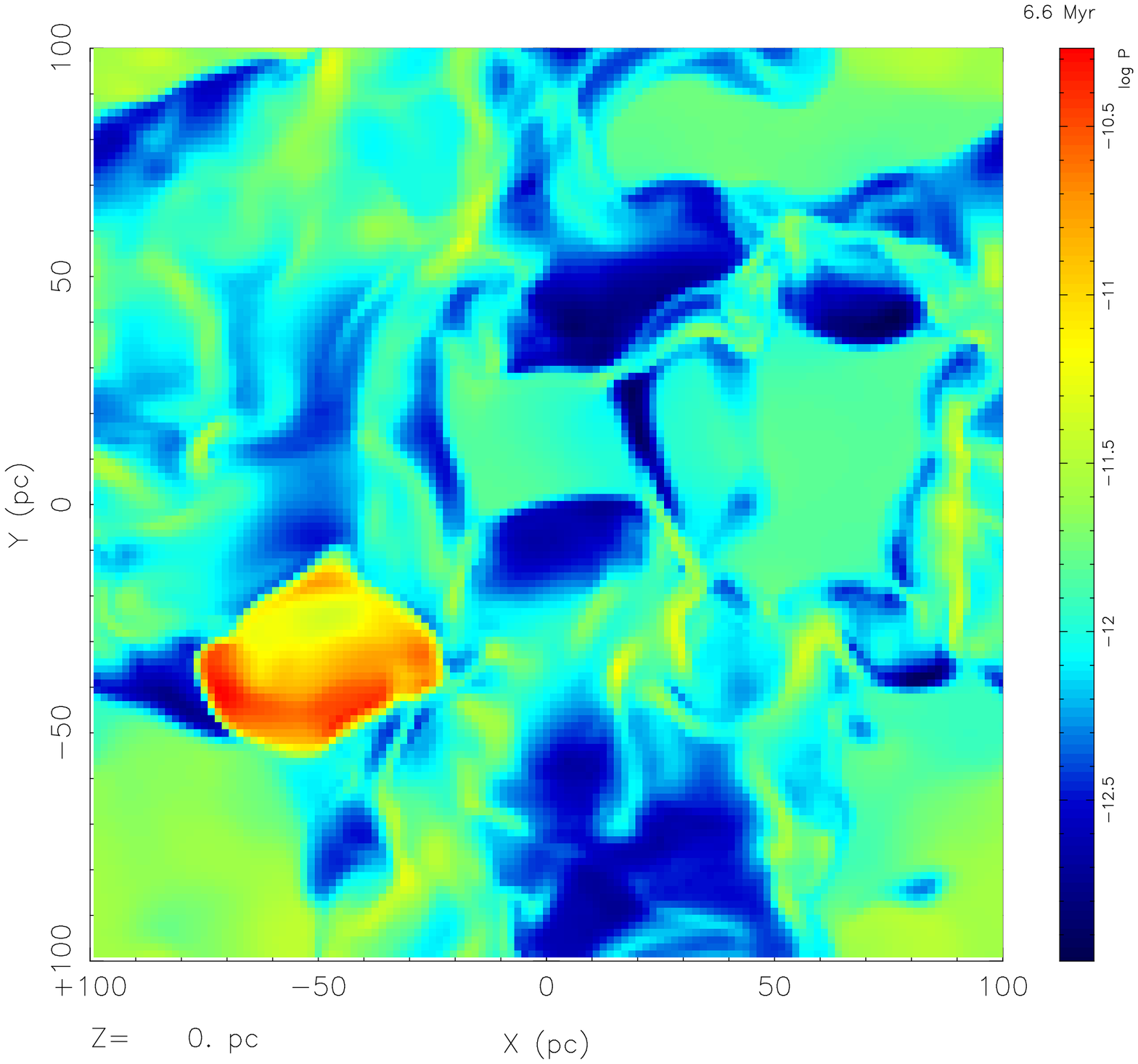}
}}
\centerline{\hbox{
\includegraphics[width=0.5\textwidth,angle=270]{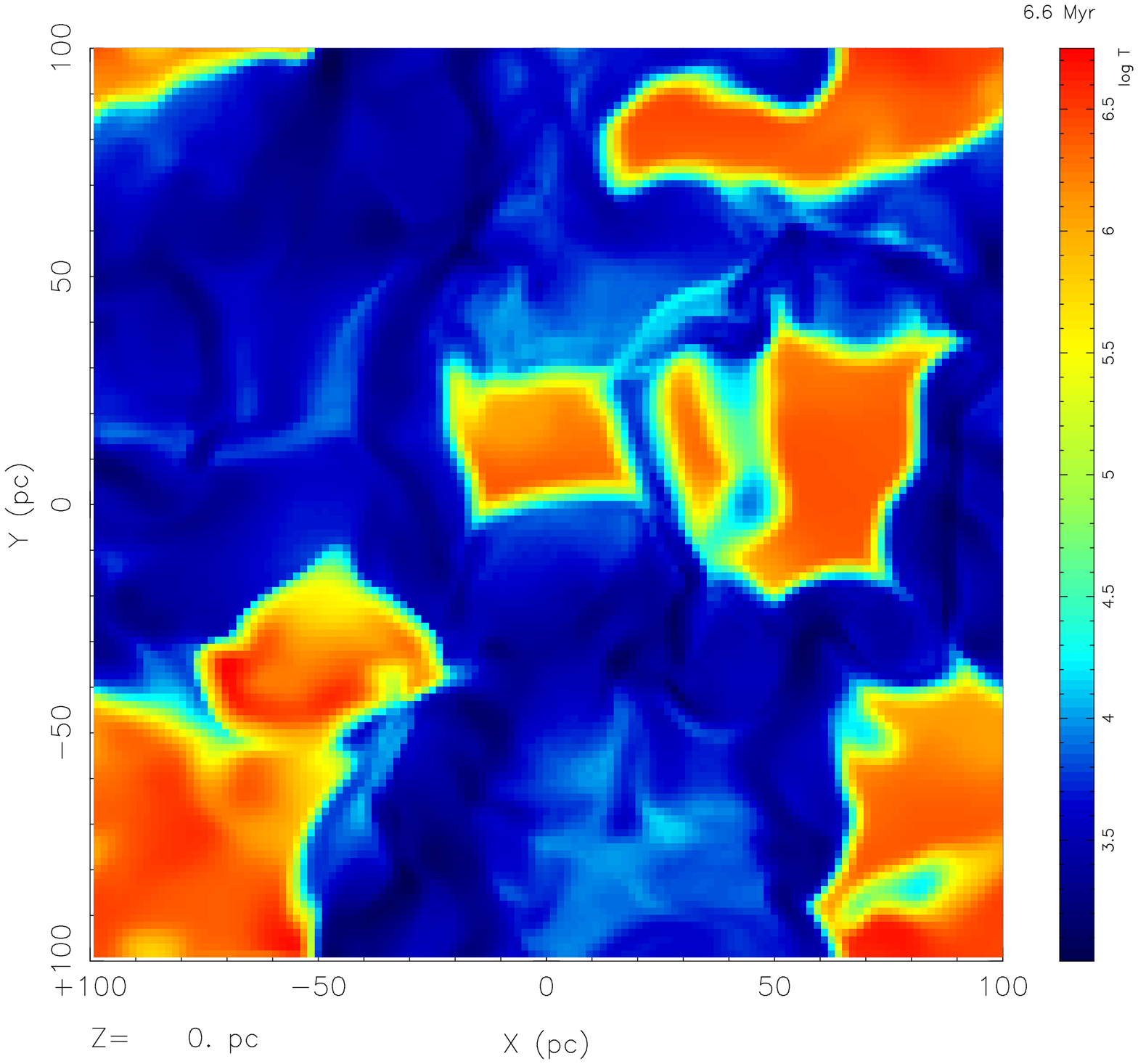}
\includegraphics[width=0.5\textwidth,angle=270]{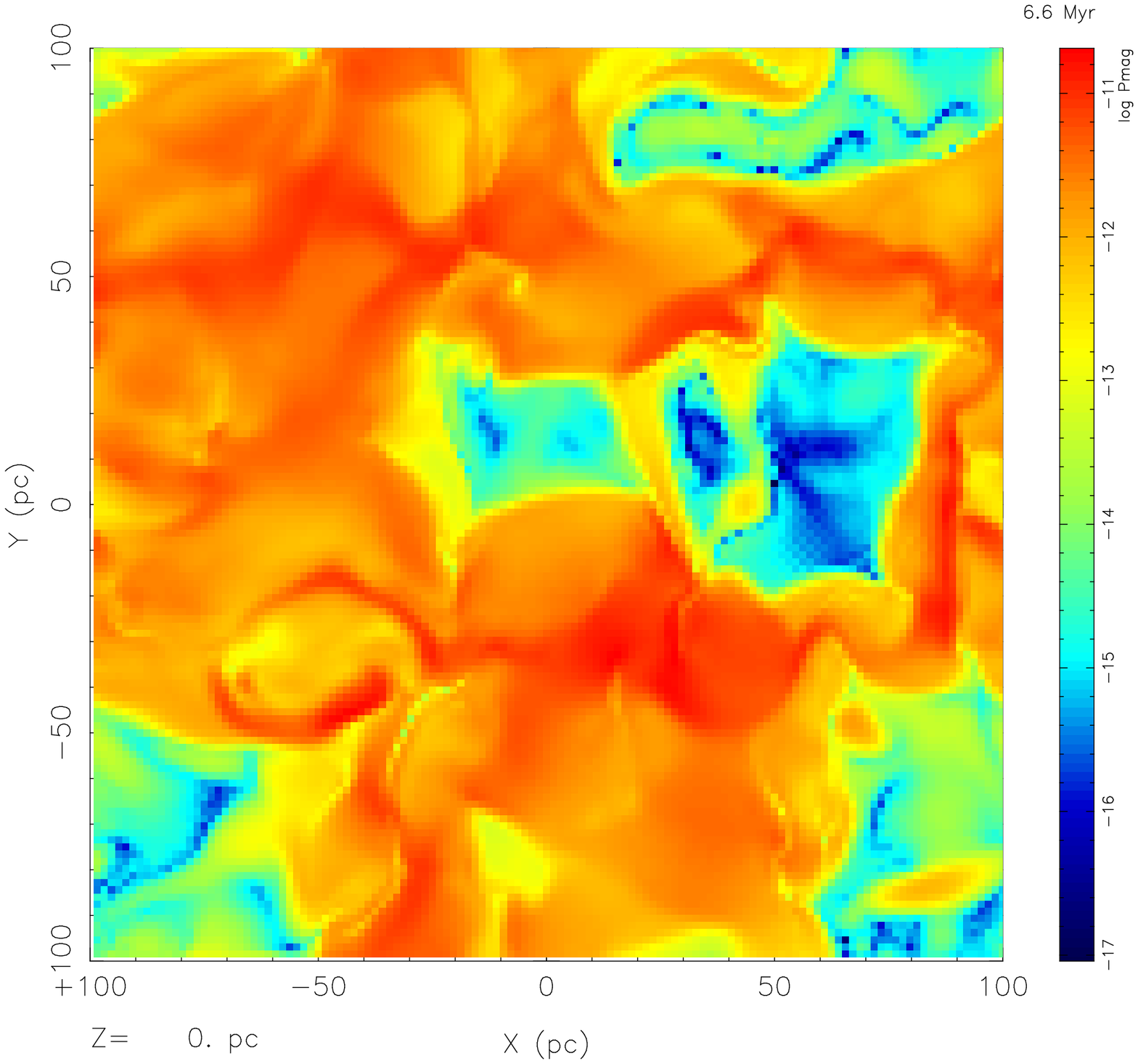}
}}
\caption[Magnetized SN-driven turbulence]{Two-dimensional
slices through the three-dimensional MHD model M2, parallel to the
magnetic field at a time of 6.6 Myr, showing density (upper left),
thermal pressure (upper right), temperature (lower left), and magnetic
pressure (lower right). Color bars indicate the scale of each
quantity. (From \cite{m02}) \label{image}}
\end{figure}

We explode SNe at a rate of one every 0.1 Myr in our box, twelve times
higher than our present Galactic rate, corresponding to a mild
starburst like M82.  The SNe are permitted to explode at random
positions. Each SN explosion dumps 10$^{51}$ erg thermal energy into a
sphere with radius 5 pc.  The evolution of the system is determined by
the energy input from SN explosions and diffuse heating and the energy
lost by radiative cooling. An example of the resulting turbulence is
shown in Figure~\ref{image}.

The first theories of the multi-phase ISM, such as Field, Goldsmith,
\& Habing~\cite{fgh69}, postulated an isobaric medium.  Since then,
multi-phase models have commonly been interpreted as being isobaric,
although McKee \& Ostriker~\cite{mo77} and Wolfire et al.~\cite{w95}
actually assume only local pressure equilibrium, not global, and McKee
\& Ostriker~\cite{mo77} considered the distribution of pressures.  In
typical multi-phase models, the heating and cooling rates of the gas
have different dependences on the temperature and density, so that the
balance between heating and cooling determines allowed temperatures
and densities for any particular pressure.  This balance can be shown
graphically in a phase diagram, showing, for example, the allowed
densities for any pressure (\cite{fgh69}; for a modern example, see
Fig.~3({\em a}) of \cite{w95}).

\begin{figure}[thbf]
\begin{center}
\includegraphics[width=1\textwidth]{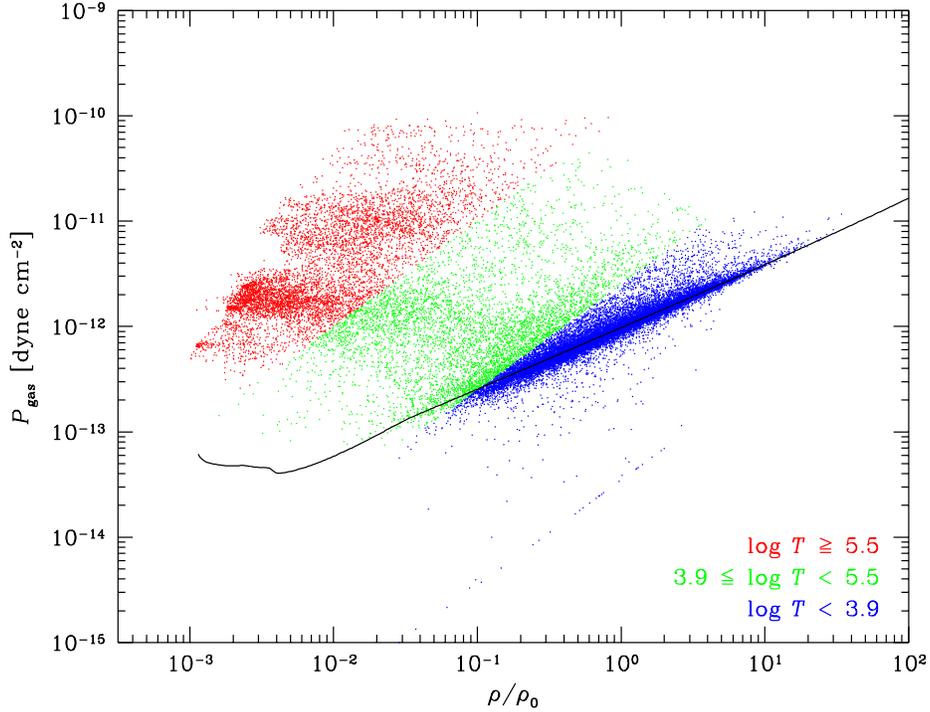}
\end{center}
\caption[Pressure vs.\ density]{Scatter plot of pressure vs.\ density at
$t=6.6$ Myrs in the MHD simulation M2, showing $32^3$ points sampled
at intervals of four points in each direction.  Note that for each
density a wide variation in pressure is seen. Cool gas with $\log T <
3.9$ is shown in blue, warm gas with $3.9 < \log T < 5.5$ in green,
and hot gas with $\log T > 5.5$ in red.  The thermal equilibrium curve
for the cooling and heating functions in this simulation is overlaid
as a black line. (The line of points at the very bottom right
corresponds to an absolute cutoff in the cooling at 100~K that was
enforced on the temperature in this model.) Note that our cooling
curve may artificially prevent much low-temperature gas from forming
in this model.  From \cite{m02} \label{press-f5}}
\end{figure}

In Figure~\ref{press-f5}, the thermal-equilibrium curve for the
heating and cooling mechanisms included is shown as a black line.
Only a single phase is predicted at high densities as our cooling
curve did not include the physically-expected unstable region at
temperatures of order $10^3$~K~\cite{w95}.  Thus, if our model
produced an isobaric medium, it would be expected to have a single
low-temperature phase in uniform density given by the point at which
the thermal-equilibrium curve crosses that pressure level.
(Effectively, we would have the hotter two of the three phases
proposed by McKee \& Ostriker~\cite{mo77}.)

The scattered points in Figure~\ref{press-f5} show the actual density
and pressure of individual zones in the model. Many zones at low
temperature do lie on the thermal equilibrium curve, but scattered all
up and down it at many different pressures and densities.
Furthermore, a substantial fraction of the gas has not had time to
reach thermal equilibrium at all after dynamical compression.  It
appears that pressures are determined dynamically, and the gas then
tries to adjust its density and temperature to reach thermal
equilibrium at that pressure.  Most gas will land on the thermal
equilibrium curve when dynamical times are long compared to heating
and cooling times.  

{\revised Even if we include the proper physics to allow multiple
phases, the behavior observed in our model} will lead to all points
within the range of pressures available along the thermal equilibrium
line being occupied, rather than the appearance of discrete phases
{\revised (also see the chapter by V\'azquez-Semadeni et al.\ in this
volume).}  Unstable regions along the thermal equilibrium
curve~\cite{g01} and off it will also be populated, as observed by
Heiles~\cite{cjh01}, but not as densely, as gas will indeed attempt to
heat or cool to a stable thermal equilibrium at its current pressure.
In particular, cold high pressure regions can be formed dynamically,
without the influence of self-gravity, perhaps giving a method for
forming molecular clouds with observed properties that are not in
hydrostatic equilibrium.

\begin{figure}[tf]
\begin{center}
\includegraphics[width=1\textwidth]{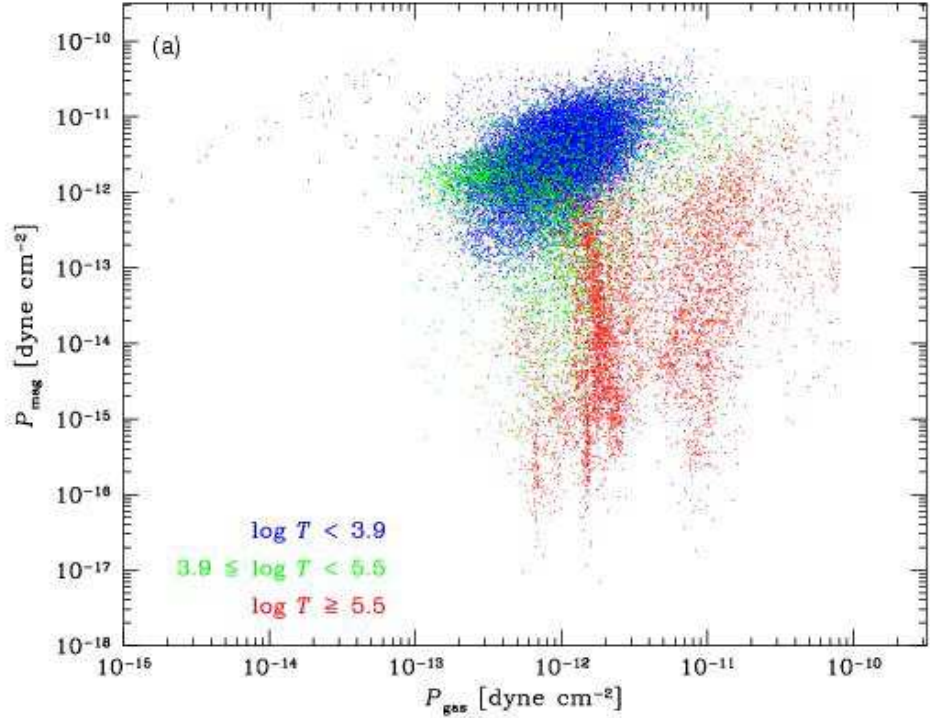}
\end{center}
\vspace{0.4in}
\caption[Magnetic pressure vs.\ thermal pressure]{Scatter plot of
magnetic vs.\ thermal pressure at $t=6.6$ Myrs in the MHD simulation.
We again plotted a subset of $32^3$ points sampled at intervals of
four points in each direction.  Note that regions of very low thermal
pressure have substantial magnetic pressures.  From \cite{m02}
\label{press-f6}}
\end{figure}

The relation between magnetic and thermal pressure is shown in
Figure~\ref{press-f6}. In this Figure,
the relative strength of thermal and magnetic pressure is shown at one
time for the magnetized simulation.  The scattering of regions at very
low thermal pressure all have substantial magnetic pressures,
demonstrating that magnetically supported regions can occur.  However,
their relative importance is rather low, as shown by the small number
of points in that regime.  Hot gas can be seen, on the other hand, to
be dominated by thermal pressure, with low magnetic pressures.

\begin{figure}[thbf]
\centerline{\hbox{
\includegraphics[width=0.5\textwidth,angle=0]{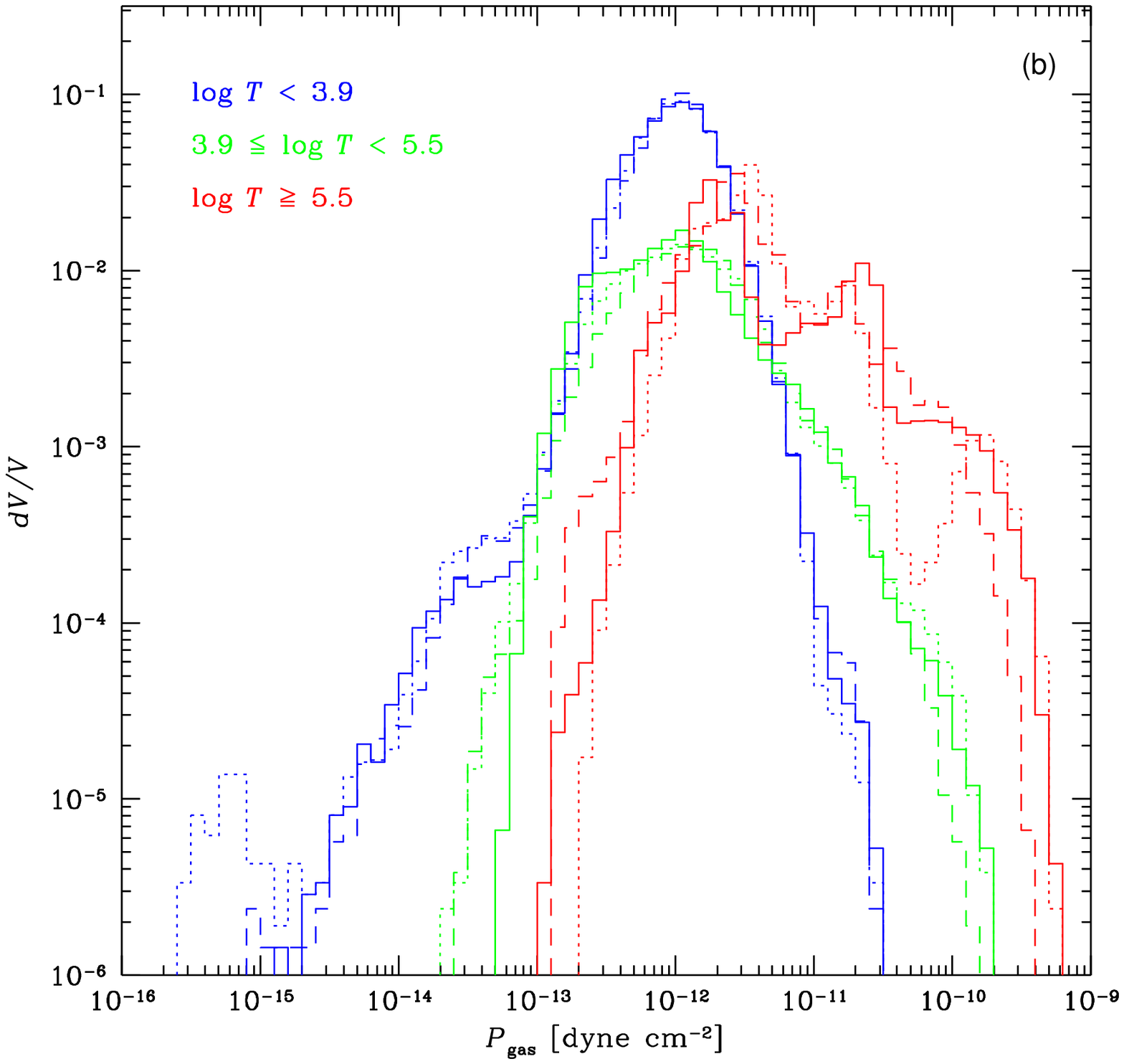}
\includegraphics[width=0.5\textwidth,angle=0]{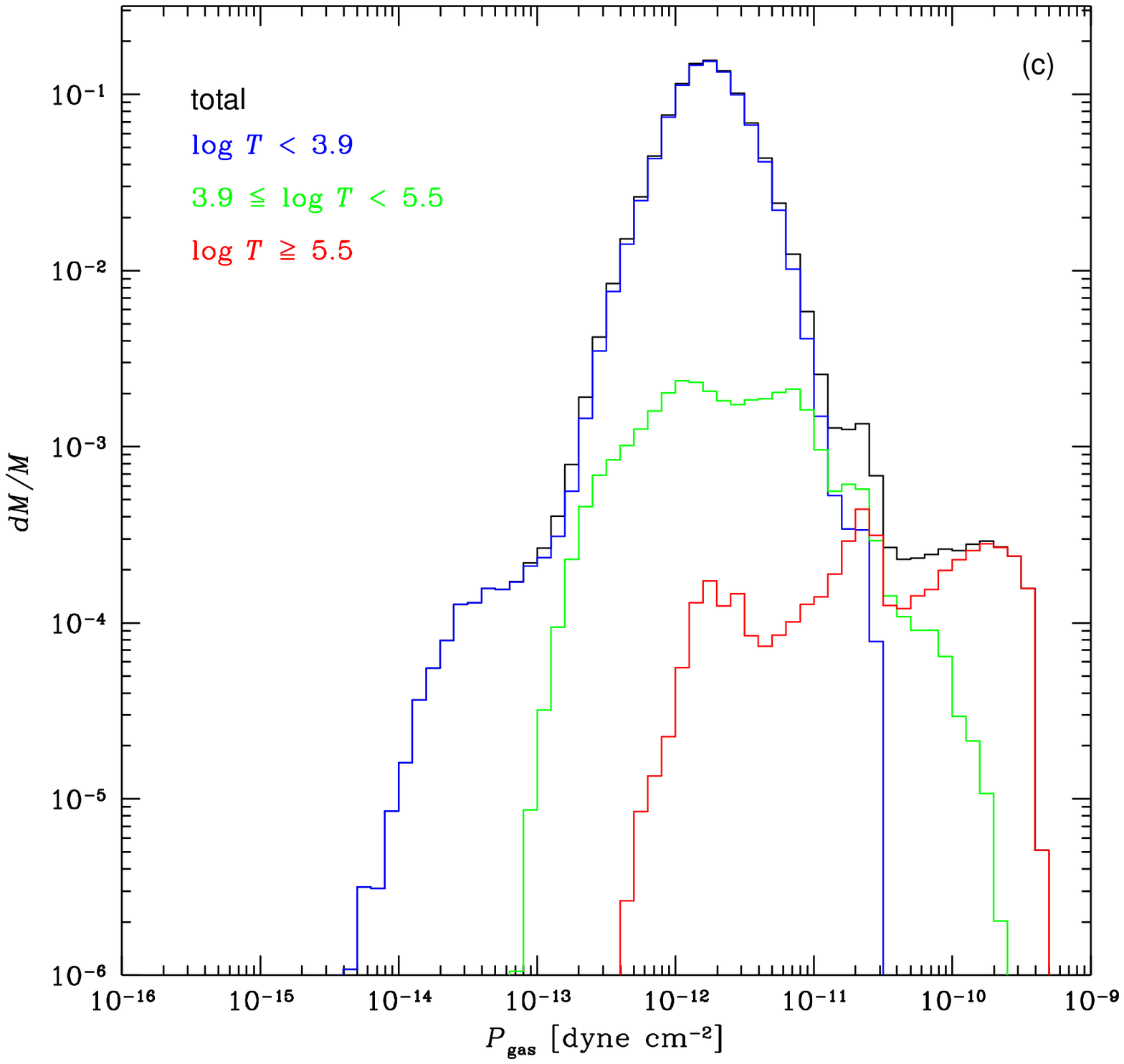}
}}
\caption[Pressure distribution]{ ({\em b}) Volume-weighted PDFs of
pressure from the MHD models for cool gas with $\log T < 3.9$ (blue),
warm gas with $3.9 < \log T < 5.5$ (green), and hot gas with $\log T >
5.5$ (red) at different times of 5.55~Myr (dashed), 6.06~Myr (dotted),
and 6.55~Myr (solid) in the 1.6~pc resolution model M2. This model has
twelve times the galactic rate of SNe, which results in a very broad
pressure distribution. ({\em c}) Mass distribution of pressure for the
MHD model M2 for the full distribution (black), and for cool gas with
$\log T < 3.9$ (blue), warm gas with $3.9 < \log T < 5.5$ (green), and
hot gas with $\log T > 5.5$ (red) at a time of 6.55~Myrs.  Most of the
mass is found in cold gas, with a broad distribution around the peak
pressure. From \cite{m02} \label{press-pdfs}}
\end{figure}

As we have shown, the supernova-driven models have broad ranges of
pressures.  We can quantify this by examining the pressure probability
density function (PDF), as shown in Figure~\ref{press-pdfs}.  In both
cases, these show roughly log-normal pressure PDFs, very unlike the
power-law distributions predicted by the analytic theory derived by
McKee \& Ostriker~\cite{mo77}.
The observed distributions rather more resemble the log-normal density
distributions suggested by Passot \&
V\'azquez-Semadeni~\cite{pv98}. Mac Low et al.~\cite{m02} show that
the cool gas can actually be modeled quite successfully with this
heuristic theory.

\section{Conclusions}

\begin{itemize}

\item Even relatively strong magnetic fields, with the field in
equipartition with the kinetic energy, cannot prevent the decay of
turbulent motions on dynamical timescales far shorter than the
observed lifetimes of molecular clouds.  The significant
kinetic energy observed in molecular cloud gas must be supplied more
or less continuously.  

\item Supersonic turbulence strong enough to globally support a
molecular cloud against collapse will usually cause {\em local}
collapse.  The turbulence establishes a complex network of interacting
shocks.  The local density enhancements in fluctuations created by
converging shock flows can be large enough to become gravitationally
unstable and collapse. The probability for this to happen, the
efficiency of the process, and the rate of continuing accretion onto
collapsed cores are strongly dependent on the driving wave length and
on the rms velocity of the turbulent flow, and thus on the driving
mechanism.

\item Interstellar clouds driven on large scales or without even
global turbulent support very rapidly form stars in clusters.  On the
contrary, in gas that is supported by turbulence, local collapse
occurs sporadically over a large time interval, forming isolated
stars.  The total star formation efficiency before the cloud dissolves
due to stellar feedback or external shocks will probably be low.
Thus, the strength and nature of the turbulence may be fully
sufficient to explain the difference between the observed isolated and
clustered modes of star formation.

\item Magnetorotational instabilities may provide a base value for the
velocity dispersion below which no galaxy will fall.  If that is
sufficient to prevent collapse, little or no star formation will
occur, producing something like a low surface brightness galaxy with
large amounts of H~{\sc i} and few stars.  In star-forming galaxies,
however, clustered and field supernova explosions, predominantly from
B~stars no longer associated with their parent gas, appear likely to
dominate the driving, raising the velocity dispersion to some 10--15
km~s$^{-1}$.

\item In a supernova-driven interstellar medium, we find a broad range
of pressures with a log-normal distribution, and a substantial
fraction of associated densities far from the thermal equilibrium
values.  This limits the predictive usefulness of phase diagrams based
on thermal equilibrium, although thermal equilibrium at the local
pressure will still be the mildly favored state.  Gas pressures appear
to be determined dynamically, and each individual parcel of gas seeks
local thermal equilibrium at the pressure imposed on it by the
turbulent flow.  Inferences that molecular clouds must be
gravitationally bound because of their high observed confinement
pressures are called into question by these results.  Regions with
densities approaching the overall densities of GMCs, and pressures an
order of magnitude above the average interstellar pressure appear in
our simulations even in the absence of self-gravity.

\end{itemize}

\vspace{0.2in} I thank {\revised the referee of this review,
E. V\'azquez-Semadeni, for a detailed and thoughtful report,} my
collaborators M. A. de Avillez, J. Ballesteros-Paredes, D. Balsara,
A. Burkert, F. Heitsch, J. Kim, R. S. Klessen, V. Ossenkopf, and
M. D. Smith for their participation in different parts of the work
reviewed here, and the organizers {\revised of the conference} for
their partial support of my attendance. This work was also partially
supported by the NSF under CAREER grant AST99-85392 and by the NASA
Astrophysical Theory Program under grant NAG5-10103. This research has
made use of NASA's Astrophysics Data System Abstract Service.

%

\end{document}